\documentclass[11pt,a4paper]{article}
\usepackage{jheppub}

\usepackage{epsfig}
\usepackage{amssymb}
\usepackage{amsmath}
\usepackage{amsfonts}
\usepackage{feynmp}
\usepackage{psfrag}

\newcommand{\be}{\begin{equation}}
\newcommand{\ee}{\end{equation}}
\newcommand{\bea}{\begin{eqnarray}}
\newcommand{\eea}{\end{eqnarray}}
\newcommand{\bi}{\begin{itemize}}
\newcommand{\ei}{\end{itemize}}
\newcommand{\ben}{\begin{enumerate}}
\newcommand{\een}{\end{enumerate}}

\newcommand{\lp}{\left(}
\newcommand{\rp}{\right)}
\newcommand{\nn}{\nonumber}

\def\gsim{\mathrel{\rlap{\lower4pt\hbox{\hskip1pt$\sim$}}
    \raise1pt\hbox{$>$}}}         %greater than or approx. symbol
\def\lsim{\mathrel{\rlap{\lower4pt\hbox{\hskip1pt$\sim$}}
    \raise1pt\hbox{$<$}}}         %less than or approx. symbol

\def \ep {\epsilon}

\title{Tree amplitudes and color decomposition \\ in broken SU(2)}

\author[1]{Liang Dai,}
\author[1]{Kirill Melnikov}
\author[1]{and Fabrizio Caola}

\affiliation[1]{Department of Physics and Astronomy, Johns Hopkins University, Baltimore, USA}

\emailAdd{ldai@pha.jhu.edu}
\emailAdd{melnikov@pha.jhu.edu}
\emailAdd{caola@pha.jhu.edu}

\abstract{We propose a color decomposition for general tree amplitudes in a 
$SU(2)$ gauge theory which is spontaneously broken via the Higgs mechanism. 
Working in the unitary gauge, we construct color-ordered amplitudes by explicitly 
presenting a set of color-ordered Feynman rules. Those primitive amplitudes are 
gauge-invariant, and they preserve perturbative unitarity in the 
high-energy limit. 
Serving as building blocks of color-dressed tree amplitudes, they allow for efficient 
evaluation of tree-level scattering amplitudes involving gauge bosons and the Higgs boson 
via the Berends-Giele recursion relations for color-ordered currents. We demonstrate the 
efficiency of this 
computational scheme by calculating on-shell amplitudes for scattering of 
five,  six  and nine $W$-bosons in the limit of vanishing Weinberg angle.
}


\begin{document}

\maketitle

\section{Introduction}

Interactions of electroweak gauge bosons 
at high energies  probe into the very nature of 
electroweak symmetry breaking.   Such interactions can, eventually, 
be studied at CERN  Large Hadron Collider (LHC). 
However, detailed investigations  of 
 the electroweak sector at  high energies require 
the development of efficient techniques to calculate amplitudes 
for scattering  processes with electroweak gauge bosons both at 
tree- and the one-loop level.  It has long been known that broken 
electroweak gauge invariance makes such perturbative computations  formidable. 

Indeed, in standard renormalizable gauges the presence of non decoupling Goldstone
bosons quickly leads to an explosion of the number of Feynman diagrams. On the other hand, 
in the unitary gauge large cancellations among the longitudinal 
 parts $p_\mu p_\nu/m_W^2$ of vector boson propagators would occur, leading to severe
numerical stability issues. 
Partly because of this, our knowledge of multi vector boson scattering is quite limited.
Tree-level results for $\gamma \gamma\to W^+ W^- ZZ$ and $\gamma \gamma \to W^+ W^- W^+ W^-$ were computed
in~\cite{Jikia:1994vz} using an optimized gauge choice. Beyond the tree-level the situation is even worse: to the best 
of our knowledge, only  the simplest case of $VV\to VV$ scattering has been studied~\cite{Denner:1995jv,Jikia:1996uu,Denner:1996ug,Denner:1997kq}.

In recent years, we have witnessed 
enormous progress in developing 
computational techniques for scattering amplitudes 
in massless gauge theories, both regular and supersymmetric (for a recent review, see e.g. the special issue~\cite{jphysa} 
and the review~\cite{MelnikovReview}). However, these techniques were mainly developed within QCD-like theories and must then be 
generalized in order for them to cope with non colored particle and with massive vector bosons. A first step in this direction
was made in~\cite{Badger:2005zh}, where processes with up to two external vector bosons were considered and in~\cite{Badger:2010eq},
where multi-photon tree-level amplitudes were studied. Another step in this direction was made in~\cite{Buchta:2010qr}, where it was shown
how to generalize the CSW construction (see~\cite{jphysa}) in order to deal with a broken gauge theory.

 A large number of on-shell computational techniques in massless gauge 
theories -- both at tree- and the one-loop level -- are based on the 
idea of color ordering. In that approach, scattering amplitudes 
are represented by sums of products of color factors and 
color-stripped objects -- the so-called color-ordered 
amplitudes. As an example, 
a useful color decomposition of $n$-gluon scattering amplitudes  
in a gauge theory with the group $SU(N_c)$ reads 
\cite{Berends1988759,Mangano1988653}
\be
\mathcal{A}^{\mathrm{tree}}_{n}\left(1,2,\cdots,n\right)=\sum_{\sigma\in S_{n}/Z_{n}}\mathrm{Tr}\left(T^{a_{\sigma(1)}}T^{a_{\sigma(2)}}\cdots T^{a_{\sigma(n)}}\right)A_{n}^{\mathrm{tree}}\left(\sigma(1),\sigma(2),\cdots,\sigma(n)\right),
\label{eq:color-decom-fund}
\ee
other useful decompositions were presented in~\cite{DelDuca200051},~\cite{MaltoniColorFlow} (see the review~\cite{MelnikovReview} for details).
Here  $A_{n}^{\mathrm{tree}}$ are color-ordered 
or primitive
 amplitudes, which only depend on the momenta and polarizations of 
external gluons.

The primitive amplitudes $A_{n}^{\mathrm{tree}}$ have many attractive properties (see e.g.~\cite{Dixon:1996wi}). 
Each primitive amplitude
receives contribution only from planar diagrams with external legs arranged in the corresponding order. 
These color-ordered diagrams can be computed by introducing a set of color-ordered Feynman rules from  
which the color degrees of freedom are removed. 
The color-stripped primitive amplitudes are gauge-invariant
and, in this sense, physical. 
Moreover, kinematic singularities of  
tree  amplitudes are closely related to their on-shell constructibility, 
as reflected by the BCFW on-shell recursion 
relation~\cite{BCFWproof,ArkaniKaplanBCFW}. 
Compared with the full color-dressed amplitude, 
primitive amplitudes, being color-ordered, 
have simpler structure of kinematic singularities; 
for this reason, they can be thought of as  basic objects for studying  
analytic properties of scattering amplitudes. 

We would like to define and work with color-ordered amplitudes to describe interactions of electroweak gauge 
bosons.  However, in a theory where gauge invariance is broken, it is not clear how to do that. 
There are multiple reasons for that, from vacuum having preferred direction in the ``color'' space, to the 
existence of color-neutral ``Higgs particle'' in the spectrum,
 which makes the concept of color ordering 
ambiguous. One option is to give up on the idea of color ordering and to generalize existing algorithms 
for calculating scattering amplitudes to make them applicable to color-dressed quantities. 
This program has been successfully carried out to 
address computation of high-multiplicity 
processes with electroweak gauge bosons  \cite{COMIX} 
and gluons \cite{ColorDressedGiele}, both 
at tree-level and beyond.

In this paper we investigate  if the concept of color ordering can be used to describe 
scattering of massive gauge bosons, in spite of the caveats pointed above. We focus on a model 
with the $SU(2)$ gauge group which is completely broken by the Higgs mechanism. We explain how to 
define color-ordered amplitudes in this model and show that those amplitudes satisfy the electroweak Ward identity 
and respect perturbative unitarity bound. 
We present explicit results for scattering amplitudes of five, 
six and nine $W$-bosons, by computing them in the 
unitary gauge  using color-ordered currents that satisfy Berends-Giele 
recursion.

The paper is organized as follows. In Section~\ref{sect1} we 
describe our model and mention some problems with arranging  the 
color decomposition of scattering amplitudes.  
 In Section~\ref{sect3} we derive  color-ordered Feynman rules and 
explain how color-ordered amplitudes are constructed.  
In Section~\ref{sect4} we prove that color-ordered currents 
satisfy electroweak Ward identity. 
In Section~\ref{sect5} we present our conclusions. 
Some results, including color-ordered Feynman rules and discussion 
of numerical computation of five- and six- and nine-$W$ scattering amplitudes 
 are relegated to the  Appendix.

\section{$SU(2)$ gauge theory, Higgs mechanism and the color decomposition}
\label{sect1}
We consider a $SU(2)$ gauge theory which is broken by the Higgs mechanism. In such a theory, three gauge fields $W^a$ are labelled by color indices $a=1,2,3$. The gauge field part of the Lagrangian reads
\be
\mathcal{L}_{gauge}=-\frac{1}{4}F_{\mu\nu}^{a}F^{a,\mu\nu},\qquad F_{\mu\nu}^{a}=\partial_{\mu}W_{\nu}^{a}-\partial_{\nu}W_{\mu}^{a}+g\varepsilon^{abc}W_{\mu}^{b}W_{\nu}^{c},
\ee
where $\varepsilon^{abc}$ is the Levi-Civita tensor. 
We use $SU(2)$ Lie algebra  generators  
$T^a = \sigma^a/\sqrt{2}$, where $\sigma^{1,2,3}$ are the Pauli 
matrices.  The orthogonality and commutation relations read   
\be
\mathrm{Tr}\left(T^{a}T^{b}\right)=\delta^{ab},\;\;\;\;
\left[T^{a},T^{b}\right]=i\sqrt{2}\varepsilon^{abc}T^{c}.
\label{eq1a}
\ee
While the above relations generalize to an arbitrary $SU(N)$ group,  
generators of the 
$SU(2)$ group enjoy an anti-commutation relation 
\be
\left\{ T^{a},T^{b}\right\} =\mathbf{1}\;\delta^{ab}, 
\label{acom}
\ee
that will play an important role in our construction. 
A completeness relation of generators in the fundamental 
representation is useful for dealing with color algebra. In the case of $SU(2)$, it reads
\be
\left(T^{a}\right)_{ij}\left(T^{a}\right)_{kl}=\delta_{il}\delta_{kj}
-\frac{1}{2}\delta_{ij}\delta_{kl}.
\label{eq_comp}
\ee

We break the gauge symmetry in the Standard Model-like way; to this end, 
we   introduce  a scalar $SU(2)$ doublet $\Phi$ and 
give it a non-vanishing vacuum expectation value
$\langle\Phi\rangle=\frac{1}{\sqrt{2}}
\left(\begin{array}{c}0\\v\end{array}\right)$.  In general, we 
 parameterize
the $SU(2)$ doublet in terms of four real scalar 
\be
\Phi=\frac{1}{\sqrt{2}}\left(\begin{array}{c}
-i\left(\phi^{1}-i\phi^{2}\right)\\
v+H+i\phi^{3}\end{array}\right).
\ee
We identify $H$ with the physical Higgs boson; 
the three fields $\phi^{a},a=1,2,3$ are Goldstone degrees 
of freedom; they are absorbed by gauge fields $W^a$, 
as they acquire equal masses $m_W=gv/2$ and obtain longitudinal modes. 
The Higgs part of the Lagrangian reads
\be
\label{eq_higgs}
\mathcal{L}_{\rm Higgs}=\left(D_{\mu}\Phi\right)^{\dagger}\left(D^{\mu}\Phi\right)+\mu^{2}\Phi^{\dagger}\Phi-\lambda\left(\Phi^{\dagger}\Phi\right)^{2},
\ee
where the covariant derivative in the fundamental representation is given by
$
D_{\mu}= \partial_{\mu}-igW_{\mu}^{a} T^{a}/\sqrt{2}.
$
This broken gauge theory can be quantized in a standard way 
by introducing a gauge-fixing term 
\be
\mathcal{L}_{g.f.}=-\frac{1}{2\xi}\left(\partial^{\mu}W_{\mu}^{a}-\xi\frac{gv}{2}\phi^{a}\right)^{2},
\ee
and the ghost Lagrangian
\be
\mathcal{L}_{\rm ghost}=\bar{u}^{a}\left[-\delta^{ab}\partial^{2}+g\varepsilon^{abc}\partial^{\mu}W_{\mu}^{c}-\xi\frac{g^{2}v}{4}\varepsilon^{abc}\phi^{c}-\xi\frac{g^{2}v}{4}\delta^{ab}H-\xi\frac{g^{2}v^{2}}{4}\delta^{ab}\right]u^{b}.
\ee
The unphysical degrees of freedom, i.e. the Goldstone fields and the ghost fields, all have masses $\sqrt{\xi} m_W$. 
In the limit  $\xi\rightarrow\infty$, which is usually referred to as the unitarity gauge, 
the unphysical degrees of freedom  decouple from the theory.
In such a gauge, intermediate states appearing in any physical 
scattering amplitude, i.e. gauge bosons $W^a$ 
and the Higgs boson $H$, are physical degrees of freedom. 
In particular, the unitary gauge 
propagator for the gauge field $W^a$ is
\be
D^{ab}_{\mu\nu}=\frac{-i\delta^{ab}}{p^2-m_W^2}\left({g_{\mu\nu}-\frac{p_{\mu}p_{\nu}}{m_W^2}}\right).
\ee
The unitary gauge deals with only  physical degrees of freedom
and, for this reason,  is particularly suitable for  unitarity-related 
tools such as on-shell recursion relation and unitarity cuts. We also 
point out that in this particular model, a global $SU(2)$ symmetry 
survives as the particle content nicely fits 
into its various representations, even though the locally gauged $SU(2)$ 
symmetry is broken. This observation 
will help  us to construct the  color decomposition in what follows.

We also mention that in this paper we restrict our discussion to self-interaction of gauge bosons and their interactions with the Higgs boson.
The self-interaction of the Higgs bosons, which can be traced back to 
the scalar potential Eq.(\ref{eq_higgs}), is not necessary to 
describe consistent interaction pattern in the gauge sector. 
Hence, the mass of the Higgs boson $m_H$ can be viewed as a free input 
parameter. Its value is not important for  ensuring the gauge invariance 
of the theory,  although it  determines the value of scattering amplitudes 
in high-energy scattering and, therefore, controls  perturbative unitarity.

It is well-known that the description of multi-particle 
scattering -- even at tree level -- becomes 
very difficult within conventional Feynman-diagrammatic approach. 
This is  especially true for gauge field  
theories where the number of Feynman diagrams 
grows factorially when the number of external particles 
increases, see Table~\ref{tab:no-feyndiag}. 
In the unitary gauge, where the number of Feynman diagrams is 
greatly reduced due to the absence of Goldstone bosons  and ghosts, 
severe cancellations occur between individual diagrams as longitudinal 
structures in propagators of gauge bosons  introduce bad scaling 
behaviors in the high energy limit. The color decomposition that we introduce 
in this Section reduces full amplitudes to simpler objects, which can be 
computed in the recursive fashion, thereby 
keeping  growth of Feynman diagrams in check and avoiding 
large numerical cancellations at intermediate 
steps.

\begin{table}
\begin{centering}
\begin{tabular}{|c|c|c|c|c|c|c|c|}
\hline 
\# of external gauge bosons & $3$ & $4$ & $5$ & $6$ & $7$ & $8$ & $9$\tabularnewline
\hline
\hline 
\# of color-dressed diagrams & $1$ & $7$ & $55$ & $730$ & $11410$ & $226765$ & $5230225$\tabularnewline
\hline
\end{tabular}
\par\end{centering}
\caption{\label{tab:no-feyndiag} Number of color-dressed tree diagrams for multi-$W$ scattering in the broken $SU(2)$ model, as generated by automation package $\mathtt{Qgraf}$\cite{qgraf}. Intermediate Higgs bosons contribute a large number of additional diagrams.}
\end{table}

A color decomposition for the $n$-gluon scattering is shown in 
Eq.(\ref{eq:color-decom-fund}).  We remind the reader that this color-decomposition 
is achieved by rewriting the structure constant $f^{abc}$ -- which enters 
Feynman rules in case of gluodynamics -- through a difference of 
traces of products of 
$SU(N_c)$ generators in the 
fundamental representation and then using the completeness 
relations to combine various traces. 

We would like to repeat the same procedure in the broken gauge theory; the immediate 
obstacle that we face is that -- in addition to the structure constants of the 
$SU(2)$ group that control self-interactions of the gauge bosons, there are 
symmetric structure constants  $\delta^{ab}$ in the coupling of the 
gauge bosons to the Higgs boson. We can  deal with the   
anti-symmetric structure  constants $\sim \varepsilon^{abc}$ 
in the standard way by writing 
\bea
\varepsilon^{abc} & = & -\frac{i}{\sqrt{2}}\left(\mathrm{Tr}\left(T^{a}T^{b}T^{c}\right)-\mathrm{Tr}\left(T^{a}T^{c}T^{b}\right)\right).
\label{eq_3}
\eea
To deal with the 
symmetric structure  constants $\delta^{ab}$, we use the fact that for 
the $SU(2)$ gauge group, they can be written  as 
anti-commutators of Lie algebra generators, Eq.(\ref{acom}).
We can employ this representation for $\delta^{ab}$ 
 to insert  it in  relevant  places  inside traces created 
by the repeated use of Eq.(\ref{eq_3}) and the completeness relation 
Eq.(\ref{eq_comp}).
We conclude that a general tree 
$W$-boson  scattering amplitude can be written as a linear combination 
of kinematic structures multiplied by traces of products 
of $SU(2)$ generators in the fundamental representation
\be
\label{eq:color-decomposition_w}
\mathcal{A}^{\mathrm{tree}}_{n}\left(1_W,2_W,\cdots,n_W\right)=
\sum_{\sigma\in S_{n}/Z_{n}}\mathrm{Tr}\left(T^{a_{\sigma(1)}}
\cdots T^{a_{\sigma(n)}}\right)
A_{n}^{\mathrm{tree}}\left(\sigma(1),\sigma(2),\cdots,\sigma(n)\right).
\ee
We note that primitive amplitudes in the above formula are defined 
to contain the gauge coupling constant in the appropriate power.  

To make use of the full power of color decomposition, it is important to understand 
how color-ordered amplitudes can be computed.  In case of pure gluodynamics, 
a powerful way to compute ordered amplitudes  is based on  Berends-Giele
recursion relations \cite{Berends1988759}.  If we want to apply a similar technique 
to compute scattering amplitudes in a broken gauge theory, we face 
the following problem: because of the existence of $WWH$ vertex, iterations of 
Berends-Giele currents for electroweak gauge bosons must involve the Higgs boson 
currents. However, since Higgs bosons are color-neutral, we face an immediate 
question of how to incorporate the neutral particles into  the color-ordering scheme. A similar issue arises if we think about using tree 
color-ordered amplitudes as building blocks 
in one-loop computations. In this case, 
unitarity cuts clearly  produce tree amplitudes 
with intermediate (multiple) Higgs particles and we need to understand how 
to define ``color-ordered'' amplitudes with Higgs particles and electroweak gauge 
bosons. 
 
We require 
that color-ordered amplitudes  receive 
contribution only from planar color-stripped diagrams with particular 
ordering of all physical external particles. Besides, we require that 
these ordered  
amplitudes satisfy electroweak Ward identity, in a similar way 
as the color-dressed amplitudes do.  This last feature -- that we will 
loosely refer to as ``gauge invariance of scattering amplitudes'' -- 
is important  for enabling 
applications of these color-ordered objects to one-loop computations. 
It turns out that for broken $SU(2)$ such color-stripped objects 
do exist. In the following Sections we construct them explicitly. 

\section{Constructing physical primitive amplitudes}
\label{sect3}

We begin by addressing the color-neutrality of the Higgs boson. 
To deal with this issue, we extend the gauge group from $SU(2)$ to $U(2)$, 
by introducing the abelian 
$U(1)$ generator $T^0=\mathbf{1}/\sqrt{2}$.  We can now consider the 
completeness relation in an $U(2)$ theory, by adding the $U(1)$ generator 
to Eq.(\ref{eq_comp}). For definiteness, we will label operators of 
$SU(2)$ with  $a,b,c$, while generators of $U(2)$ will be labeled 
with  ${\tilde a}$, etc. The tilded indices run from $0$ to $3$ while the 
untilded ones run from $1$ to $3$.  For the $U(2)$ group, we still have 
the  commutation relation 
\be
\left[T^{\tilde a},T^{ \tilde b}\right]=i \sqrt{2} f^{\tilde a \tilde b \tilde c }T^{\tilde c},
\label{eq1}
\ee
where $f^{\tilde a \tilde b \tilde c}$ vanishes if any of the indices is zero and 
$f^{\tilde a \tilde b \tilde c} = \varepsilon^{\tilde a \tilde b \tilde c}$ 
otherwise. 
In addition, the simplified completeness relation is valid
\be
\left(T^{\tilde a}\right)_{ij}\left(T^{\tilde a}\right)_{kl}
=\delta_{il}\delta_{kj}.
\label{eq_comp_u2}
\ee
On the other hand, the anticommutation relation Eq.(\ref{acom}) 
requires care since 
it becomes invalid for a generic choice of $U(2)$ generators. 

We can now extend the particle content of the theory 
by promoting the gauge bosons 
and the Higgs boson to full $U(2)$ multiplets. This implies that we introduce 
the Higgs triplet $H^a$, $a=1,2,3$, in addition to the regular $SU(2)$ Higgs 
boson  that (in this notation) is denoted as
$H^0$, and the $U(1)$ gauge boson $W^0$.  The interactions 
between these particles are controlled by $U(2)$ Feynman rules.  In the gauge 
boson sector, we obtain those rules by writing 
\be
\varepsilon^{abc} \to  \varepsilon^{\tilde a \tilde b \tilde c} 
= -\frac{i}{\sqrt{2}}  \left[{\rm Tr} \left ( T^{\tilde a} T^{\tilde b} T^{\tilde c} \right ) 
 - {\rm Tr} \left ( T^{\tilde b} T^{\tilde a} T^{\tilde c} \right )\right]. 
\ee
It is then obvious that with this extension of the Feynman rules, the $U(1)$ gauge 
bosons completely decouple  from the gauge sector of the 
theory  although it is useful to have them, 
to prove the color decomposition in a straightforward way.

In the Higgs sector,  we need to extend 
the interactions between $W$-bosons and the Higgs boson. Again, we want to 
make this extension in such a way, that the decoupling of unphysical particles 
is obvious. Recall that, eventually, we are interested in computing multi-$W$ and 
multi-Higgs scattering amplitudes where all external states are taken to be physical.
To this end, we write $W^{\tilde a}W^{\tilde b}H^{\tilde c}$ vertex as
\bea
\label{eq1w}
\nonumber\\
\parbox{3cm}{
  \begin{fmffile}{WWH}
    \begin{fmfgraph*}(80,60)
      \fmfleft{v1,v2}
      \fmfright{v3}
      \fmflabel{${\tilde a},\mu$}{v1}
      \fmflabel{${\tilde b},\nu$}{v2}
      \fmflabel{${\tilde c}$}{v3}
      \fmf{boson}{v1,i1}
      \fmf{boson}{v2,i1}
      \fmf{dashes}{v3,i1}
      \fmfdot{i1}
    \end{fmfgraph*}
  \end{fmffile}
}
&&\;\;\;\;=\frac{igm_{W} g^{\mu\nu}}{\sqrt{2}}\left(\mathrm{Tr}\left(
T^{\tilde a}T^{\tilde b}T^{\tilde c}\right)+\mathrm{Tr}\left(T^{\tilde b}T^{\tilde a}T^{\tilde c}\right)\right).
\\ \nonumber 
\eea

It is easy to understand that this equation leads to decoupling of the interaction between unphysical 
Higgses $H^{a}$, $a=1,2,3$ 
and physical gauge bosons. Indeed, in this  case  
${\rm Tr}(T^aT^bT^c) \sim \varepsilon^{abc}$, 
so the sum of the two traces vanishes.  The 
non-vanishing contribution requires that either $\tilde a= \tilde b= \tilde c=0$, 
which  gives an  interaction of a physical $H$ with two unphysical $W$-bosons 
or that one of $\tilde a, \tilde b, \tilde c$ 
is zero and the other two are not. This latter 
case contains an interaction of a physical $W$ with unphysical Higgs and 
unphysical $W$, as well as the interaction of a physical Higgs 
with two physical $W$-bosons.  An important feature of the above vertex 
is that 
unphysical particles always appear in pairs; this will be a 
crucial element 
for understanding their decoupling from physical amplitudes.

Similarly, the $WWHH$ vertex 
can be generalized in the following way 
\bea
\label{eq2w}
\nonumber\\
\parbox{3cm}{
  \begin{fmffile}{WWHH}
    \begin{fmfgraph*}(80,60)
      \fmfleft{v1,v2}
      \fmfright{v4,v3}
      \fmflabel{${ \tilde a},\mu$}{v1}
      \fmflabel{${\tilde b},\nu$}{v2}
      \fmflabel{${\tilde c}$}{v3}
      \fmflabel{${\tilde d}$}{v4}
      \fmf{boson}{v1,i1}
      \fmf{boson}{v2,i1}
      \fmf{dashes}{v3,i1}
      \fmf{dashes}{v4,i1}
      \fmfdot{i1}
    \end{fmfgraph*}
  \end{fmffile}
}
=&&
\frac{ig^{2}}{4}g^{\mu\nu} 
\left ( {\rm Tr} \left ( T^{\tilde a} T^{\tilde b} T^{\tilde e} \right ) 
+  {\rm Tr} \left ( T^{\tilde b} T^{\tilde a} T^{\tilde e} \right ) \right ) 
 \\
&&\times 
\left ( {\rm Tr} \left ( T^{\tilde c} T^{\tilde d} T^{\tilde e} \right ) 
+  {\rm Tr} \left ( T^{\tilde d} T^{\tilde c} T^{\tilde e} \right ) \right ). 
\nonumber
\eea
The right hand side produces a variety of vertices that involve both physical 
and unphysical particles; again, 
the unphysical particles appear in pairs. 

By repeated use of the completeness relation for Lie algebra generators
Eq.(\ref{eq_comp_u2}),  we combine individual traces into traces 
of products of $T$-matrices that correspond to color-states 
of all external particles, including the Higgs boson. The relevant 
color-stripped Feynman rules  are given in  Appendix~\ref{sec:appendix_fr}.
We emphasize that the $U(2)$ Feynman rules imply that unphysical 
particles can be produced in pairs only; therefore, 
if external particles are physical,  the unphysical particles 
automatically decouple from full tree amplitude, in spite of contributing 
to color-ordered ones. This decoupling is identical to how ghosts  in QCD
or super-partners in supersymmetric QCD 
 do not  contribute to scattering  amplitudes of regular quarks and gluons 
  at tree level. We therefore conclude that scattering 
amplitudes can be represented in the following way 
\be
\label{eq:color-decomposition}
\mathcal{A}^{\mathrm{tree}}_{n}\left(1_X,2_X,...n_X \right ) = 
\sum_{\sigma\in S_{n}/Z_{n}}\mathrm{Tr}\left(T^{{\tilde a}_{\sigma(1)}}T^{{\tilde a}_{\sigma(2)}}
\cdots T^{{\tilde a}_{\sigma(n)}}\right)
A_{n}^{\mathrm{tree}}\left(\sigma(1),\cdots,\sigma(n)\right),
\ee
where $X^{\tilde a}$ is a generic notation for the Higgs boson and $W$ bosons, 
and  color-ordered amplitudes $A_{n}$ are obtained from 
the color-ordered Feynman rules. 

As we pointed out already, we would like to construct 
the color-ordered amplitudes that 
satisfy electroweak Ward identity that connects 
matrix elements of ``gauge'' and ``Goldstone'' currents  
\bea
\label{pic:gauge-inv-wc}
\nonumber\\
\frac{k^{\mu}}{m_W}\cdot
\parbox{3cm}{
  \begin{fmffile}{gaugeinv}
    \begin{fmfgraph*}(80,60)
      \fmfleft{v1}
      \fmfright{v2,v3,v4,v5}
      \fmf{boson,label=$\mu\quad k$,tension=3}{v1,i1}
      \fmf{boson}{v2,i1}
      \fmf{boson}{v3,i1}
      \fmf{boson}{v4,i1}
      \fmf{boson}{v5,i1}
      \fmfblob{.5w}{i1}
    \end{fmfgraph*}
  \end{fmffile}
}
=-i\cdot
\parbox{4cm}{
  \begin{fmffile}{gaugeinv2}
    \begin{fmfgraph*}(80,60)
      \fmfleft{v1}
      \fmfright{v2,v3,v4,v5}
      \fmf{dots,label=$\phi$,tension=3}{v1,i1}
      \fmf{boson}{v2,i1}
      \fmf{boson}{v3,i1}
      \fmf{boson}{v4,i1}
      \fmf{boson}{v5,i1}
      \fmfblob{.5w}{i1}
    \end{fmfgraph*}
  \end{fmffile}
}.
\eea
This  relation between matrix elements of the two currents should be 
valid in any gauge, including the unitary one. 

For the purpose of computing color-ordered amplitudes, 
we can define the color-stripped currents, where all external particles
-- including the Higgs bosons --  
are ordered, in full analogy with QCD.  
Similar to ordered amplitudes,  these currents 
can be computed as sums  of all color-ordered diagrams,
using a set of color-ordered Feynman rules. The $n$-particle 
partial amplitude is obtained by computing the scalar product of a 
$(n-1)$-point current 
with the polarization vector of the $n$-th particle  and taking the on-shell 
limit\footnote{We do not include the external propagator of $n$-th particle 
into   the definition of the current.}
\be
A_{n}^{\mathrm{tree}}\left(1,2,\cdots,n-1,n\right)=
\lim_{k_1^2 \to m_W^2}
\epsilon_{\mu}\left(k_{1}\right)W^{\mu}\left(2,\cdots,n-1,n\right).
\ee

We envision that ordered amplitudes can, eventually, be used in one-loop 
computations based on generalized unitarity \cite{MelnikovReview}.  To 
enable computations in the unitary gauge, it is crucial that 
ordered amplitudes satisfy the electroweak Ward identity, in the sense of 
Eq.(\ref{pic:gauge-inv-wc}) since this relation allows, formally, 
to start a calculation in the Feynman gauge and then argue that, after 
taking the unitarity cuts, contributions of 
unphysical $W$-polarization and  contribution of 
the Goldstone boson cancel out exactly, leaving out the unitary gauge result. 
 Therefore, we require that  
Eq.(\ref{pic:gauge-inv-wc}) holds for color-ordered currents 
\be
\label{eq:gauge-inv-nc}
\frac{k_{1}^{\mu}}{m_{W}}W_{\mu}\left(2,\cdots,n-1,n\right)=
-iG\left(2,\cdots,n-1,n\right).
\ee

Finally, we require that partial amplitudes are 
perturbative-unitary. By this we mean that amplitudes for 
gauge-boson scattering approach  a constant in the limit of infinitely 
large center-of-mass energy. Explicitly, 
\be
\label{eq:unitarity-nc}
A^{\mathrm{tree}}_{n}\left(1,2,\cdots,n\right)=\mathrm{constant}+\mathcal{O}\left(\frac{m_{W}^{2}}{s_{ij}},\frac{m_{H}^{2}}{s_{ij}}\right),\qquad 
s_{ij}\rightarrow \infty ,
\ee
where $1,2,\cdots,n$ can be gauge bosons of any polarization or Higgs bosons, and $s_{ij}=\left(k_i+k_j\right)^2$ where $i,j$ represents 
any two of the external particles. Empirically, 
we find that, after enforcing gauge 
invariance, perturbative unitarity works out automatically. 

Since we plan to use unitary gauge for the computation  
of color-ordered amplitudes, we do not need to discuss 
interaction vertices  where Goldstone bosons appear. However,  
we need such vertices to check the electroweak Ward identity. 
Having in mind the unitary gauge, we only 
require a vertex with a single Goldstone boson $\phi H W$.
Indeed, vertices 
with larger number of Goldstone bosons lead to diagrams where  Goldstone bosons appear as 
 internal particles; such diagrams decouple in the unitary gauge, 
due to the infinitely large mass of the Goldstone boson.
We will take 
\bea
\label{eq:gauge-inv-wc1}
\nonumber\\
\parbox{4cm}{
  \begin{fmffile}{WGH2}
    \begin{fmfgraph*}(60,60)
      \fmfleft{v1}
      \fmfright{v3,v2}
      \fmflabel{${\tilde a},\mu$}{v1}
      \fmflabel{${\tilde b},\nu$}{v2}
      \fmflabel{${\tilde c}$}{v3}
      \fmf{boson}{v1,i1}
      \fmf{dots,label=$p_1$}{v3,i1}
      \fmf{dashes,label=$p_2$}{v2,i1}
      \fmfdot{i1}
    \end{fmfgraph*}
  \end{fmffile}
}
\hspace*{-2cm}=\frac{g}{2\sqrt{2}}\left(p_1-p_2\right)^{\mu}
\\ \nonumber 
\eea
as the color-stripped Feynman rule for the interaction of the Goldstone boson with physical degrees of freedom 
and check if this is sufficient to maintain gauge invariance. 

\begin{figure}
\centering
\epsfig{file=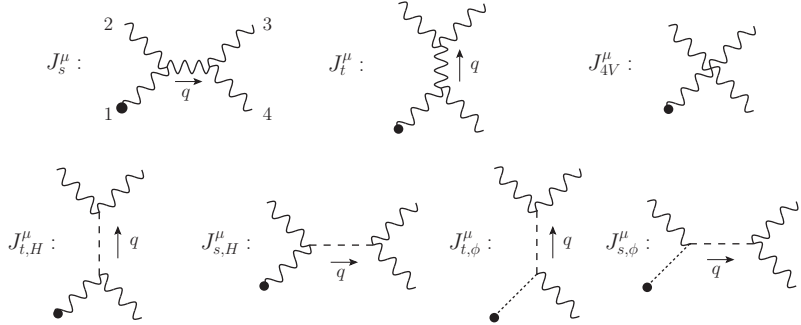,width=0.7\textwidth}
\caption{Currents for the $0\to WWWW$ Ward identity.}\label{fig:WWWW_curr}
\end{figure}

We are now in position to start checking the electroweak Ward 
identity for 
color-ordered amplitudes. 
We begin with the ordered amplitude that describes scattering of 
four $W$-bosons $ 0 \to W_1(p_1) + W_2(p_2) + W_3(p_3) + W_4(p_4)$.
The corresponding diagrams 
are shown in Fig.~\ref{fig:WWWW_curr}. 
The vertices that contribute to the description 
of the $W$-boson scattering all follow from the color-stripped version of vertices that naturally arise in the 
unitary gauge, see Eqs.(\ref{eq1w}).  The right hand side of 
the Ward identity Eq.(\ref{eq:gauge-inv-nc})
is even simpler and receives contributions from the color-ordered vertex 
shown in Eq.(\ref{eq:gauge-inv-wc1}).

Because of the symmetry of the problem, we only have to check the Ward 
identity  with respect to the momentum of one gauge boson.  
We choose the gauge boson 
with momentum $p_1$ for this purpose.  
We 
write the scattering amplitude as 
\be
{\cal M} = 
\epsilon_{1\mu} \left ( J_s^\mu+J_t^\mu+J_{4W}^\mu 
+J_{s,H}^\mu 
+J_{t,H}^\mu 
\right ),
\ee
where the currents are defined in Fig.~\ref{fig:WWWW_curr}. 
We begin by considering all 
diagrams without the intermediate Higgs boson.
We consider first the $s-$channel diagram that contributes to 
$J_s^\mu$.   We obtain
\be
\begin{split} 
J^\mu_s =& \lp \frac{i g}{\sqrt2}\rp^2
\left[(p_1-p_2)^\alpha \ep_2^\mu + (p_2-q)^\mu \ep_2^\alpha - 2 p_1\cdot \ep_2 g^{\mu\alpha}\right]\frac{i}{q^2-m_W^2}
\\
& \times 
\lp -g_{\alpha\beta} + \frac{q_\alpha q_\beta}{m_W^2}\rp
\left[(p_3-p_4)^\beta \ep_3\cdot \ep_4 + 2 p_4\cdot \ep_3 \ep_4^\beta - 2 p_3\cdot \ep_4 \ep_3^\beta\right],
\end{split}
\ee
with $q = -p_1-p_2$. To contract $J_s$ with $p_1$, we note that
\be
p_{1,\mu}\cdot\left[(p_1-p_2)^\alpha \ep_2^\mu + (p_2-q)^\mu \ep_2^\alpha - 2 p_1\cdot \ep_2 g^{\mu\alpha}\right]
=q^\alpha p_1\cdot \ep_2 + (q^2-m_W^2) \ep_2^\alpha
\ee
and
\be
\left[q^\alpha p_1\cdot \ep_2 + (q^2-m_W^2) \ep_2^\alpha\right]\lp -g_{\alpha\beta}+\frac{q_\alpha q_\beta}{m_W^2}\rp=
-(q^2-m_W^2) \ep_{2,\beta}.
\ee
Putting everything together we obtain
\be
p_1\cdot J_s = \frac{i g^2}{2}\left[(p_3-p_4)\cdot \ep_2 (\ep_3\cdot\ep_4) + 2 p_4\cdot \ep_3 (\ep_2\cdot\ep_4)- 2 p_3\cdot \ep_4 (\ep_2\cdot\ep_3)\right].
\ee
Contribution of the 
$t-$channel diagram $J_t^\mu$ is easily obtained from 
the $J_s^\mu$ by interchanging 2 and 4. The result is 
\be
p_1\cdot J_t = \frac{i g^2}{2}\left[(p_3-p_2)\cdot \ep_4 (\ep_2\cdot\ep_3) + 2 p_2\cdot \ep_3 (\ep_2\cdot\ep_4)- 2 p_3\cdot \ep_2 (\ep_3\cdot\ep_4)\right].
\ee
The $4W$-vertex gives a contribution
\be
p_1\cdot J_{4W} = ig^2\left[
p_1\cdot \ep_3 (\ep_2\cdot\ep_4) - \frac{1}{2} p_1\cdot \ep_2 (\ep_3\cdot\ep_4) - \frac{1}{2} p_1\cdot\ep_4 (\ep_2\cdot \ep_3)
\right].
\ee
Putting together the ``pure-gauge'' contributions, we find
\begin{align}
p_1\cdot (J_s+J_t+J_{4W}) =& \frac{ig^2}{2}\left[\ep_2\cdot\ep_3 (-2p_3 + p_3-p_2-p_1)\cdot \ep_4 \right.\nn\\
&+\left.\ep_2\cdot\ep_4 (2p_4+2 p_2 + 2 p_1)\cdot \ep_3 + \ep_3\cdot\ep_4 (p_3-p_4-2p_3-p_1)\cdot\ep_2\right] \nn\\
=&\frac{ig^2}{2}\left[(\ep_2\cdot\ep_3) p_4\cdot \ep_4 - 2 (\ep_2\cdot\ep_4) p_3\cdot\ep_3 + (\ep_3\cdot\ep_4)p_2\cdot\ep_2\right] \nn\\
=&0.
\end{align}

We now consider diagrams with the intermediate Higgs boson.
In the $s-$channel we have
\be
p_1\cdot J_{s,H} = 
\left[\frac{ig}{\sqrt2} m_W p_1\cdot\ep_2\right]
\frac{i}{q^2-m_H^2}
\left[\frac{ig}{\sqrt2} m_W \ep_3\cdot\ep_4\right]
\ee
with $q=-p_1-p_2$. If we use the physical condition $p_2\cdot\ep_2=0$,
 we can write 
$p_1\cdot \ep_2 = -(q-p_1)\cdot\ep_2/2$ and 
\be
p_1\cdot J_{s,H} = -i m_W
\left[\frac{g}{2\sqrt2}(q-p_1)\cdot\ep_2\right]
\frac{i}{q^2-m_H^2}
\left[\frac{ig}{\sqrt2} m_W \ep_3\cdot\ep_4\right].
\ee
Because of the Feynman rule shown in Eq.(\ref{eq:gauge-inv-wc1}), 
this result is in exactly $-i m_W J_{s,\phi}$.   
The same result clearly holds also for $J_{t,H}$ and $J_{t,\phi}$.
 Hence, we conclude that in case of $W$-scattering, 
the Ward identity holds and it works out in the following way:
it holds diagram by diagram for Higgs 
exchanges while the sum of diagrams that only involve gauge bosons
is  transverse on its own in the unitarity gauge.  
This result is summarized in Fig.~\ref{fig:WWWW}:   the four-$W$ 
ordered amplitude  satisfies the relevant Ward identity, 
without the need for new interaction vertices.
We note that a tight relation 
between Higgs 
exchanges and pure gauge scattering diagrams comes from the requirement 
that color-ordered amplitudes are unitary. We have checked that perturbative unitarity  holds for $4W$ scattering.

\begin{figure}
\centering
\epsfig{file=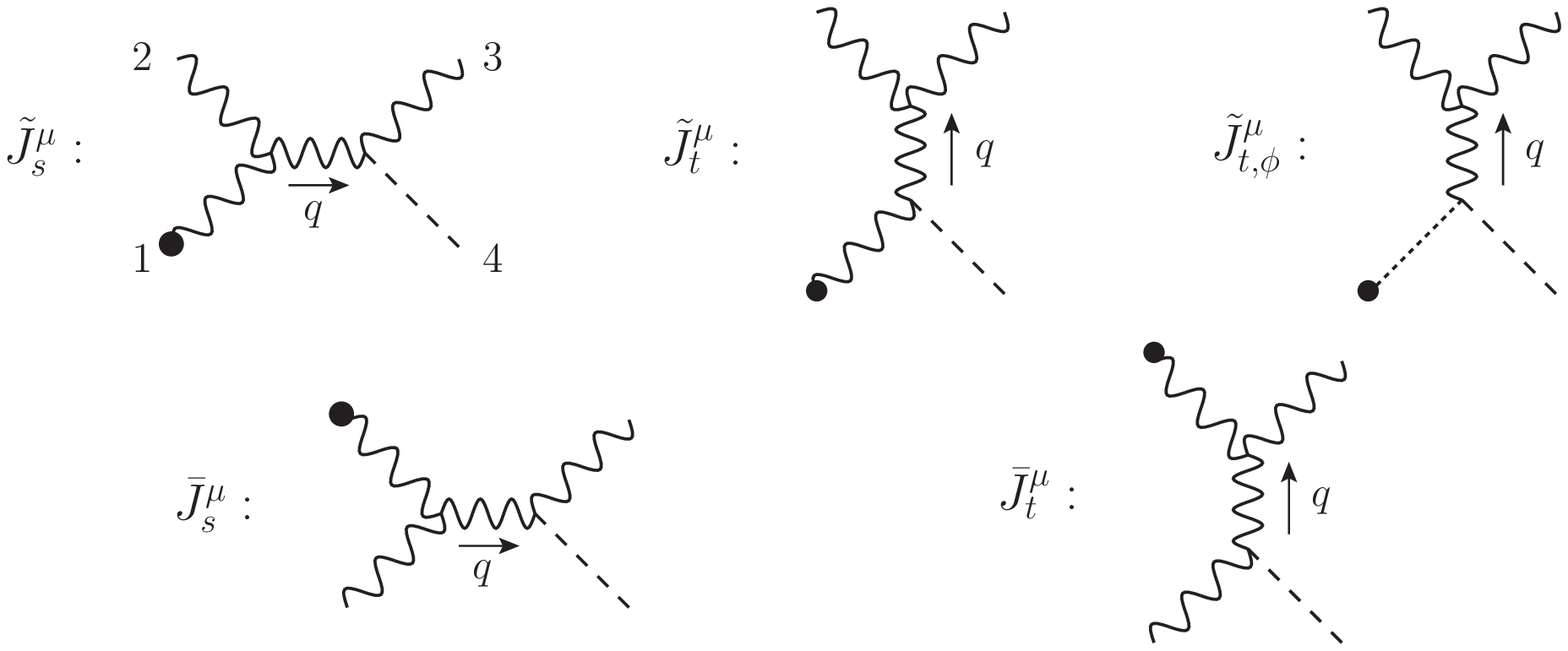, width=0.7\textwidth}
\caption{Currents for the $0\to WWWH$ Ward identity.}\label{fig:WWWH_curr}
\end{figure}

As the next step, we consider the $0 \to W_1(p_1) + W_2(p_2)+ W_3(p_3) 
+ H_4(p_4) $  ordered 
scattering amplitude. This is no longer a fully symmetric case, 
and we have to check three different Ward identities separately, 
one for each $W$ leg.  However, given the fact that ordered amplitudes 
are cyclic-symmetric, only two cases are independent. The relevant currents are
shown in Fig~\ref{fig:WWWH_curr}.

We begin by checking the Ward identity with respect to the $W_1$ leg. 
We write the scattering amplitude as 
\be
{\cal M} = \epsilon_{1\mu} \left ( {\tilde J}_s^\mu + {\tilde J}_t^\mu \right ).
\ee
Using partial  results presented in the discussion of  four-$W$ scattering 
amplitude,  
it is straightforward to compute the $s-$channel contribution
\be
p_1\cdot {\tilde J}_s = \left[\frac{ig}{\sqrt2}(m_W^2-q^2) \ep_2^\beta\right]
\frac{i}{q^2-m_W^2}
\left[
\frac{ig}{\sqrt2}m_W \ep_{3,\beta}
\right]=\frac{ig^2}{2} m_W \ep_2\cdot \ep_3.
\label{eq_js}
\ee
For the $t-$channel contribution we can write
\be
p_1\cdot {\tilde J}_t 
= \left[\frac{ig}{\sqrt2}m_W p_1^\alpha\right]\frac{i}{q^2-m_W^2}
\lp -g_{\alpha\beta} + \frac{q_\alpha q_\beta}{m_W^2}\rp V_3^\beta(-q,p_2,p_3),
\ee
where $V_3$ is the all-outgoing three-boson vertex. 
Writing  $p_1 = -(q+p_4-p_1)/2$, we
obtain
\begin{align}
 p_1\cdot \tilde J_t =& -i m_W\left[\frac{g}{2 \sqrt2}(p_4-p_1)^\alpha\right]\frac{i}{q^2-m_W^2}
\lp -g_{\alpha\beta} + \frac{q_\alpha q_\beta}{m_W^2}\rp V_3^\beta(-q,p_2,p_3)\nn\\
& -i m_W\left[\frac{g}{2 \sqrt2}q^\alpha\right]\frac{i}{q^2-m_W^2}
\lp -g_{\alpha\beta} + \frac{q_\alpha q_\beta}{m_W^2}\rp V_3^\beta(-q,p_2,p_3). 
\end{align}
Since $q\cdot V_3(-q,p_2,p_3) = 0$, 
the second line vanishes and the first term coincides with $-i m_W \tilde J_{t,\phi}$.
Hence, if we put everything together, we  obtain the {\it violation} of the Ward identity -- 
the divergence of the $s$-channel contribution 
does not match any term on the right hand side of the Ward identity.
Explicitly, we obtain  
\be
p_1\cdot W(2,3,4) = - i m_W G(2,3,4) + \frac{ig^2}{2} m_W \ep_2\cdot \ep_3.
\ee
At the color-dressed level the offending term cancels between 
$s$- and $u$-channel  contributions but,  if we want 
{\it ordered amplitudes} to satisfy the Ward identity, 
we need to introduce additional vertices. The simplest one to introduce  
to enforce the Ward identity  is a 
local $\phi WW H$ vertex. We can take the corresponding color-ordered 
vertices  to be 
\bea
\nonumber \\
\parbox{4cm}{
  \begin{fmffile}{2WGH2}
    \begin{fmfgraph*}(80,60)
      %\fmfpen{thick}
      \fmfleft{v1,v2}
      \fmfright{v4,v3}
      \fmflabel{$\mu$}{v2}
      \fmflabel{$\nu$}{v3}
      \fmf{boson}{v2,i1}
      \fmf{boson}{v3,i1}
      \fmf{dots}{v1,i1}
      \fmf{dashes}{v4,i1}
      \fmfdot{i1}
    \end{fmfgraph*}
  \end{fmffile}
}
\hspace*{-1cm}=\frac{g^2}{4}g^{\mu\nu},
\qquad \qquad
\parbox{4cm}{
  \begin{fmffile}{2WGH1}
    \begin{fmfgraph*}(80,60)
      %\fmfpen{thick}
      \fmfleft{v1,v2}
      \fmfright{v4,v3}
      \fmflabel{$\mu$}{v1}
      \fmflabel{$\nu$}{v2}
      \fmf{boson}{v1,i1}
      \fmf{boson}{v2,i1}
      \fmf{dots}{v3,i1}
      \fmf{dashes}{v4,i1}
      \fmfdot{i1}
    \end{fmfgraph*}
  \end{fmffile}
}
&&\hspace*{-1cm}=-\frac{g^2}{4}g^{\mu\nu}.
\\ \nonumber
\eea
The color-dressed version of this vertex is constructed in such a way that, when the sum 
over all colors is taken, this vertex vanishes. 
This is important for ensuring that 
this vertex does not contribute to color-dressed amplitudes.

We are now in position to check the Ward identity  
for the $W_2$ boson. We write the amplitude as 
\be
{\cal M} = \epsilon_{2\mu} \left ( {\bar J}_s^\mu + {\bar J}_t^\mu \right ).
\ee
%We stress that -- in principle, 
%we must consider the potential contribution 
%from $\phi WW H$ vertex introduced above.
We start with the $s-$channel current. 
We can use the result in Eq.(\ref{eq_js}), after $1  \leftrightarrow 2$ flip. 
The flip gives a minus sign and we obtain 
\be
p_2\cdot J_s = -\frac{ig^2}{2}m_W \ep_1\cdot \ep_3.
\ee
The same is true for the $t-$channel diagram. We just have to take the result above and exchange 1 with 3. In this case the $VVV$ vertex
picks up a minus sign, while the $VVH$ vertex is unchanged. We are left then with an overall minus sign, with the result
\be
p_2\cdot J_t = +\frac{ig^2}{2}m_W \ep_3\cdot \ep_1.
\ee
We see that the Ward identity is satisfied without additional vertices. 
Hence, we must forbid the 
four-particle ordered vertex $W\phi W H$,
in spite of the existence of the ordered vertex  $\phi WW H$.

\begin{figure}
\centering
\epsfig{file=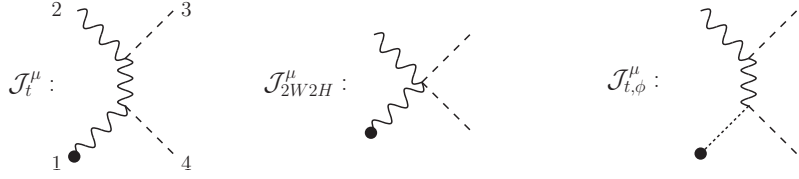,width=0.7\textwidth}
\caption{Currents for the $0\to WWHH$ Ward identity.}\label{fig:WWHH_curr}
\end{figure}

Finally, we have to consider amplitudes with two $W$-bosons and 
two Higgs bosons, 
$0 \to W(p_1)+ W(p_2) + H (p_3) + H(p_4)$. 
Because of symmetry, 
we only have to check the Ward identity with respect to one of the 
vector bosons; we choose the $W_1$. The relevant currents are shown in 
Fig.~\ref{fig:WWHH_curr}.
 We write the amplitude as 
\be
{\cal M} = \epsilon_{1\mu} \left ( {\cal J}_s^{\mu} + {\cal J}_t^{\mu} 
+ {\cal J}_{2W2H}^{\mu} \right ).
\ee
We start with the $t-$channel contribution 
\begin{align}
p_1\cdot {\cal J}_t =& 
-i m_W\left[\frac{g}{2 \sqrt2}(p_4-p_1)^\alpha\right]\frac{i}{q^2-m_W^2}
\lp -g_{\alpha\beta} + \frac{q_\alpha q_\beta}{m_W^2}\rp 
\left[\frac{ig}{\sqrt2}m_W \ep_2^\beta\right]\nn\\
&-i m_W\left[\frac{g}{2 \sqrt2}q^\alpha\right]\frac{i}{q^2-m_W^2}
\lp -g_{\alpha\beta} + \frac{q_\alpha q_\beta}{m_W^2}\rp \left[\frac{ig}{\sqrt2}m_W \ep_2^\beta\right].
\end{align}
The first line cancels with the single term on 
the right-hand of the Ward identity,  while the second one gives
\be
-i m_W\left[\frac{g}{2 \sqrt2}q^\alpha\right]\frac{i}{q^2-m_W^2}
\lp -g_{\alpha\beta} + \frac{q_\alpha q_\beta}{m_W^2}\rp \left[\frac{ig}{\sqrt2}m_W \ep_2^\beta\right]=
\frac{ig^2}{4} q\cdot \ep_2
\ee
The $WWHH$ vertex gives a contribution
\be
p_1\cdot {\cal J}_{2W2H} = \frac{ig^2}{4} p_1\cdot\ep_2, 
\ee
so that -- once we put everything together -- 
we find a mismatch in the Ward identity
\be
p_1\cdot W(2,3,4) = - i m_W G(2,3,4) + \frac{ig^2}{4} (p_1+q)\cdot \ep_2.
\ee
To fix the last term, we modify the $W(2,3,4)$ current by making use of the fact that 
the $s$-channel current mediated by the exchange of the $W$-boson is propagator-free, when 
contracted with $p_1$.   Since we need $HH$ final state, we introduce the ordered 
$WHH$ interaction  vertex
\bea
\label{eq:gauge-inv-wc}
\parbox{3cm}{
  \begin{fmffile}{WHH_cl}
    \begin{fmfgraph*}(60,60)
      \fmfleft{v1}
      \fmfright{v3,v2}
      \fmflabel{$\tilde a, \mu$}{v1}
      \fmf{boson}{v1,i1}
      \fmf{dashes,label=$p_1$}{v2,i1}
      \fmf{dashes,label=$p_2$}{v3,i1}
      \fmfdot{i1}
    \end{fmfgraph*}
  \end{fmffile}
}
&&\hspace*{-1cm}=\frac{g}{2}\left(p_1-p_2\right)^{\mu}.\\
\nonumber 
\eea
To ensure that this vertex does not contribute to color-dressed amplitudes 
with external physical particles, we assign the $f^{\tilde a \tilde b \tilde c}$ color factor 
to it. As a result, unphysical particles are produced in pairs by this
vertex and, therefore, do not contribute to amplitudes with physical external 
particles.  Note, however, that the new $WHH$ interaction vertex leads to 
additional contributions to $WWWH$ amplitude that we already considered 
and found to satisfy the Ward identity without it. Therefore, we have 
to make sure that the addition of the new vertex does not destroy the Ward 
identity. A simple computation shows that the Ward identity 
for $WWWH$ amplitude remains valid even after the addition of diagrams 
with new vertices. The full Ward identities with additional couplings are shown in 
Appendix~\ref{sec:appendix_wy}.

\begin{figure}[t]
\centering
\includegraphics[scale=0.6]{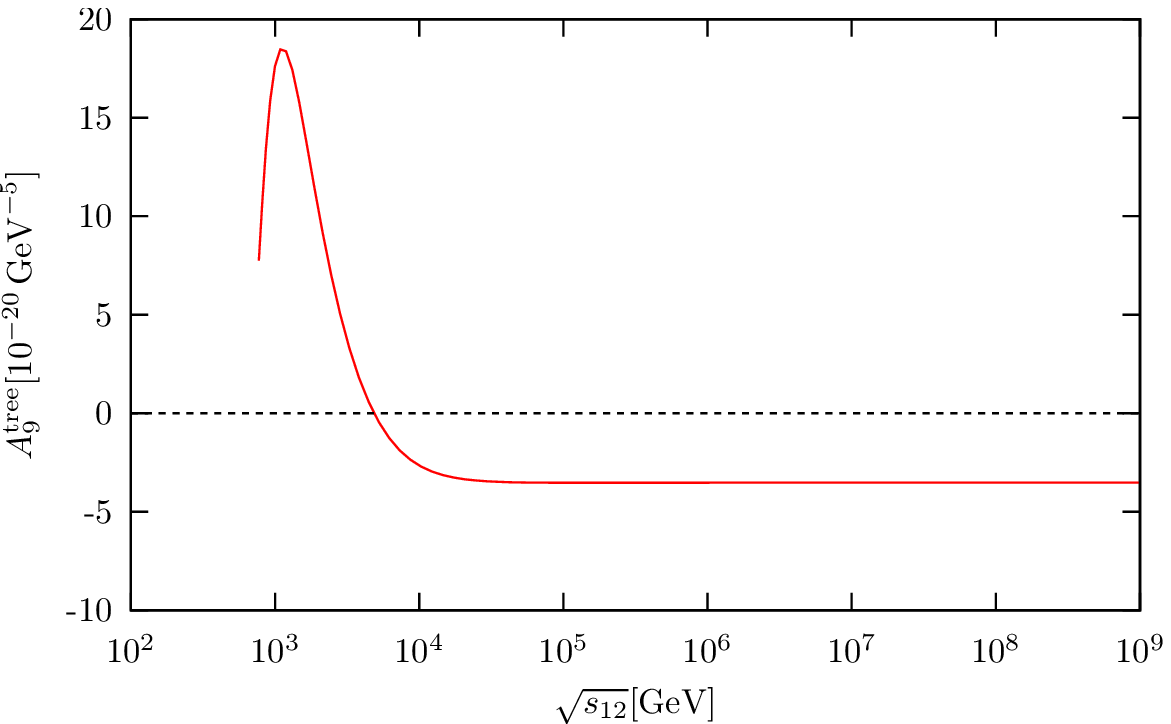}\;\;\;\;\;
\includegraphics[scale=0.6]{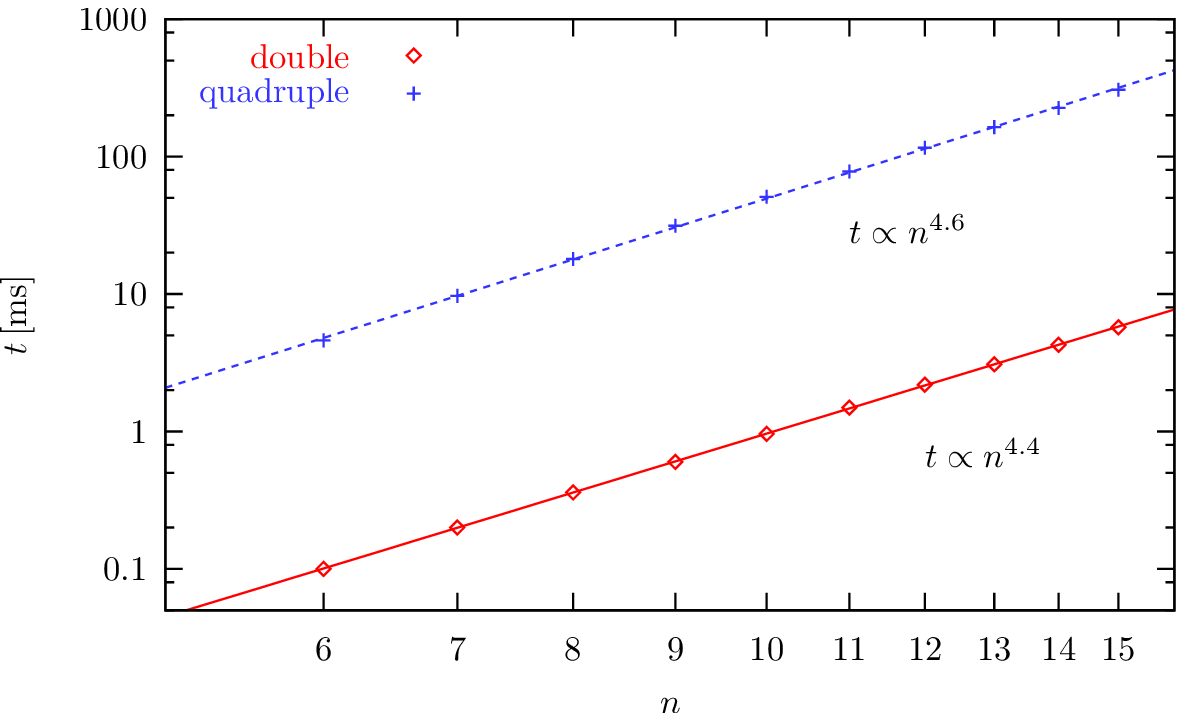}
\caption{\label{fig:time} 
Left pane: a typical color-ordered amplitude as a function of the center-of-mass energy  
clearly shows behavior consistent with perturbative unitarity. Right pane: time required to compute a typical partial amplitude for $W$-scattering under double or quadruple precision.
}
\end{figure}

\begin{figure}[t]
\centering
\includegraphics[scale=0.6]{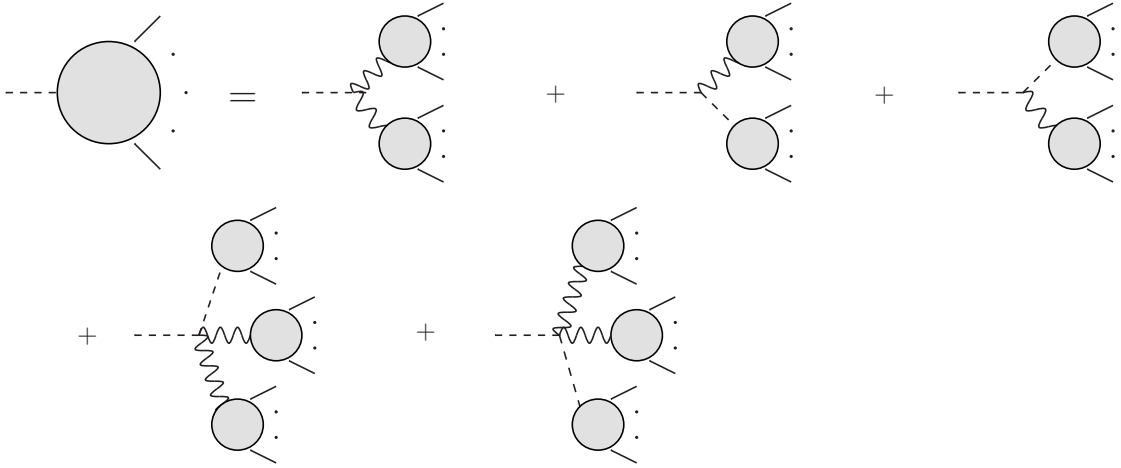}
\caption{\label{fig:recursion-relation-H} Recursion relation for color-ordered $n$-point function coupled to $H$. }
\end{figure}

\begin{figure}[t]
\centering
\includegraphics[scale=0.6]{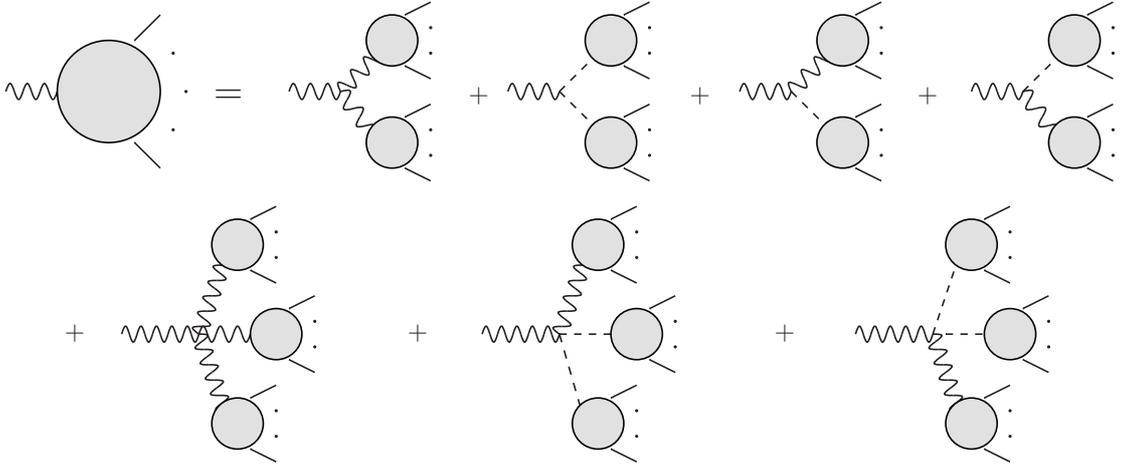}
\caption{\label{fig:recursion-relation-W} Recursion relation for color-ordered $n$-point current coupled to $W$.}
\end{figure}

\begin{figure}[t]
\centering
\includegraphics[scale=0.6]{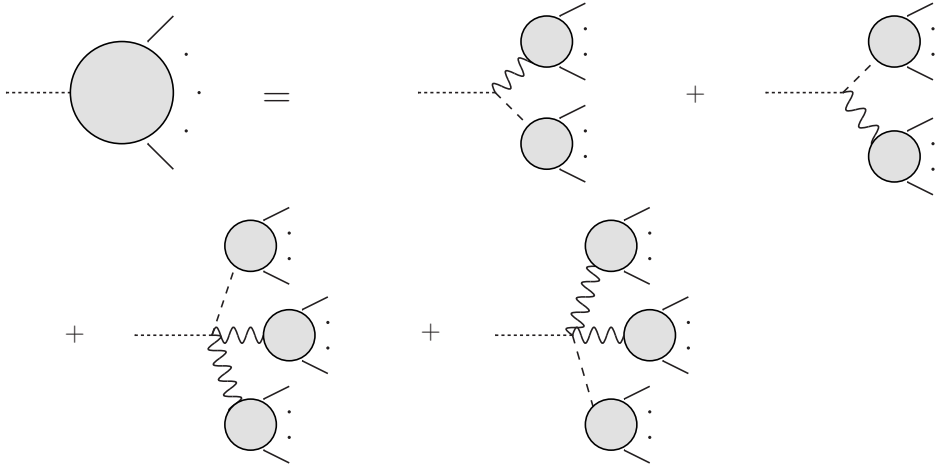}
\caption{\label{fig:recursion-relation-G} Recursion relation for color-ordered $n$-point function coupled to $G$}
\end{figure}

In fact, the color-ordered Feynman rules that appeared in the 
discussion of four-particle scattering amplitudes 
are sufficient to define tree-level 
color-ordered amplitudes for arbitrary multiplicities  of external states. 
Given the color-ordered Feynman rules, partial amplitude $A^{\mathrm{tree}}_n\left(1,\cdots,n\right)$ 
can be computed in an extremely efficient way using the Berends-Giele recursion relations for 
off-shell currents. The Berends-Giele recursion relations for 
the Higgs boson current, the gauge boson current,   and for the Goldstone boson current  are
shown in Figs.\ref{fig:recursion-relation-H},\ref{fig:recursion-relation-W},\ref{fig:recursion-relation-G}. These color-ordered currents 
satisfy coupled recursion relations, 
which resemble the recursion relations in QCD~\cite{Berends1988759} but are more complicated. 
%In the diagrammatic representation, the summation marks imply summation 
%over all possible splitting (but ordered) via Feynman vertices.

Working in an unitary gauge, we checked numerically that partial amplitudes satisfy the 
Ward identity  Eq.(\ref{eq:gauge-inv-nc}) and are unitary  Eq.(\ref{eq:unitarity-nc}) 
for up to 9 external particles. We have also checked that full amplitude obtained by computing 
color-ordered amplitudes and assembling them into a full amplitude using 
Eq.(\ref{eq:color-decomposition}) agrees with the full scattering amplitude computed with Feynman 
diagrams. 

The left pane in Fig.~\ref{fig:time} shows the dependence of the ordered amplitude 
for 9-$W$ scattering as a function of the collision energy. It is clear from that Figure 
that the ordered amplitude approaches the constant limit at high-energy, consistent with 
perturbative unitarity. We have checked that this behavior is typical for other ordered 
multi-particle  amplitudes. Note that zeroes of the ordered amplitude
 also exist, but we checked that they do not correspond to zeroes of the full amplitude.
In the next Section, we prove that the color-ordered currents 
that can be constructed using color-ordered Feynman rules satisfy 
electroweak Ward identity Eq.(\ref{eq:gauge-inv-nc}).

The right pane in Fig.~\ref{fig:time} shows the time required to compute a single color-ordered amplitude.
We have checked that -- for $n$-point amplitude the time scales like $n^{4.4}$ under double precision, roughly independent 
of the type  and polarizations of external particles. This time scaling is similar to what has 
been achieved in computations of pure gluon amplitudes, see e.g. Ref.~\cite{Giele:2008bc}. 
The roughly $n^4$ scaling can be understood as follows: (1) to calculate $(n-1)$-point currents 
one needs to calculate $1$-point, $2$-point,...,$(n-2)$-point currents, requiring $\mathcal{O}\left(n\right)$ recursions; 
(2) there are $\mathcal{O}\left(n\right)$ $(n-1)$-point ordered currents to calculate; 
(3) to calculate each $(n-1)$-point ordered current via recursion relation, 
the maximum number of ways to split is $\mathcal{O}\left(n^2\right)$, corresponding to 4-point vertices. 
One would expect improved scaling as $n^3$ if 4-point vertices are traded for 3-point vertices by introducing auxiliary fields.
 Similar conclusion has also been achieved  in the study of 
color-dressed recursions \cite{ColorDressedGiele}. We also compared the time required 
for calculating comparable processes using 
our recursive code and $\mathtt{MadGraph}$5~\cite{MadGraph5}; the results of the comparison 
are shown in Table~\ref{tab:time-cost-compare}.
We found the recursive approach much more efficient than the traditional Feynman 
diagrammatic approach, especially for large number of external gauge bosons. 

\begin{table}
\begin{centering}
\begin{tabular}{|c|c|c|c|c|}
\cline{1-4} 
\multicolumn{2}{|c|}{$\mathtt{MadGraph}5$} & \multicolumn{2}{c|}{recursive} & \multicolumn{1}{c}{}\tabularnewline
\hline 
process & time & process & time & ratio \tabularnewline
\hline 
$WW\rightarrow4W$\;\;\;\;\;\;\; & $0.026\,\mathrm{s}$ & $WW\rightarrow4W$ & $0.006\,\mathrm{s}$ & $4.3$\tabularnewline
\hline 
$WW\rightarrow4W+Z$ & $6.66\,\mathrm{s}$ & $WW\rightarrow5W$ & $0.072\,\mathrm{s}$ & $92.5$\tabularnewline
\hline
\end{tabular}
\par\end{centering}
\caption{\label{tab:time-cost-compare} Efficiency comparison between the recursive method and $\mathtt{MadGraph}$. 
Computation  was  performed on the same computer, in double precision.    
We studied comparable, but not identical, processes with the same number 
of external particles and similar number of Feynman diagrams.  The $\mathtt{MadGraph}$ 
calculation refers to full weak-boson scattering in the 
Standard Model, while  the recursive computation refers to 
the broken $SU(2)$ model that we consider in this paper.}
\end{table}

\section{Proof of electroweak Ward identity for arbitrary multiplicity}
\label{sect4}

In this Section 
we present the recursive proof
of the Ward identity Eq.(\ref{eq:gauge-inv-nc}) 
for  ordered  tree amplitudes of arbitrary multiplicity, 
generalizing the  discussion in Ref.\cite{Berends1988759} to the case 
of currents in broken gauge theory.   
To facilitate the proof, we introduce  some compact notations. 
We  consider $n$-point off-shell currents, where the 
$n$ on-shell physical particles have
outgoing momenta $k^{\mu}_i,\;i=1,2,\cdots,n$ and are physically ordered. 
We use $\tilde{W}_a$,$\tilde{H}_a$ and $\tilde{G}_a$ to 
denote off-shell currents 
coupled to gauge bosons $W$, the Higgs boson $H$ and the Goldstone 
$G$, respectively. Note that for the sake of compactness, we do not 
display Lorentz index of the gauge current. We also choose not to multiply 
the currents by an off-shell propagator.

The subscript in currents ${\tilde W}_a({\tilde H}_a)$, 
$a=1,2,3,\cdots$,  is used to indicate how 
many currents the original current  has been divided  
into, and their relative ordering; summation over all possible 
partitions  is  implicitly assumed.
For example a term ${\tilde H}_1\left({\tilde W}_2\cdot {\tilde W}_3\right)$ 
is the shorthand notation for 
\be
\sum_{i=1}^{n-2}\sum_{j=i+1}^{n-1}{\tilde H}\left(1,i\right)\left({\tilde W}\left(i+1,j\right)\cdot {\tilde W}\left(j+1,n\right)\right),
\ee
and a term ${\tilde H}_1{\tilde H}_2\left(p_2\cdot {\tilde W}_3\right){\tilde H}_4$ is the shorthand notation for 
\be
\sum_{i=1}^{n-3}\sum_{j=i+1}^{n-2}\sum_{k=j+1}^{n-1}{\tilde H}\left(1,i\right){\tilde H}\left(i+1,j\right)\left(p\left(i+1,j\right)\cdot {\tilde W}\left(j+1,k\right)\right){\tilde H}\left(k+1,n\right).
\ee
where the momentum sum $p(i,j)$ is defined as
\be
p\left(i,j\right)=\sum_{m=i}^{j}k_{m}.
\ee

We also denote by $W_a$ and $H_a$ the currents where 
the propagator of the off-shell leg is multiplied in 
(since Goldstone boson  only appears as external current in 
the unitary gauge, 
there is no need to define such an object also for $G$). These currents 
read
\be
W_{a}=\frac{-i}{p_{a}^{2}-m_{W}^{2}}\left(\tilde{W}_{a}-\frac{p_{a}\cdot\tilde{W}_{a}}{m_{W}^{2}}p_{a}\right),
\;\;\;\;\;
H_{a}=\frac{i}{p_{a}^{2}-m_{H}^{2}}\tilde{H}_{a}.
\ee

We want to show that 
\be
p_{a}\cdot\tilde{W}_{a}=-im_{W}\tilde{G}_{a},
\qquad p_{a}\cdot W_{a}=\frac{1}{m_{W}}\tilde{G}_{a}.
\ee
We prove the Ward identity by induction. 
The induction starts with one-particle  
gauge current, which is simply the polarization vector 
of the relevant particle. Note that a single-particle 
Goldstone current vanishes since all external 
particles are physical. The Ward identity for 
single-particle currents then follows trivially.

Furthermore, we introduce a useful notation to describe 
three- and four-point gauge vertices
\bea
&&W_{1}\left(-2p_{1}
-p_{2}\right)\cdot W_{2}+W_{2}\left(p_{1}+2p_{2}\right)\cdot 
W_{1}+\left(p_{1}-p_{2}\right)\left(W_{1}\cdot W_{2}\right) =
\left[W_{1},W_{2}\right]
,\\
&&2W_{2}\left(W_{1}\cdot W_{3}\right)-W_{1}\left(W_{2}\cdot W_{3}\right)-W_{3}\left(W_{1}\cdot W_{2}\right) =
\left\{ W_{1},W_{2},W_{3}\right\}.
\eea
Using  this  notation, Berends-Giele recursion relations 
can be written in a compact way. For example, the recursion 
relation for the gauge current reads 
\be
\tilde{W} =\frac{ig}{\sqrt{2}}\left[W_{1},W_{2}\right]+\frac{ig^{2}}{2}\left\{W_{1},W_{2},W_{3}\right\}
+\cdots\cdots
\label{eq_rr}
\ee
The advantage of this notation is that recursion relations can be 
re-inserted into the right-hand-side of the recursion 
relation in  a compact way. For example
\be
\tilde{H}_{1}\left(\tilde{W}_{2}\cdot\tilde{W}_{3}\right)=\frac{ig}{\sqrt{2}}\tilde{H}_{1}\left[W_{2},W_{3}\right]\cdot\tilde{W}_{4}+\frac{ig^{2}}{2}\tilde{H}_{1}\left\{ W_{2},W_{3},W_{4}\right\} \cdot\tilde{W}_{5}+\cdots\cdots,
\ee
where in the right hand side we wrote the recursion relation for $\tilde{W}_2$ and re-named the currents 
following the convention that they are numbered according to their clock-wise appearance.

We will prove by induction that 
the Ward identity 
\be
p \cdot\tilde{W} - (-im_W)\tilde{G}=0
\label{eq:gauge-inv-eq}
\ee
holds  for any multiplicity of the external 
particles.
First, we consider the gauge current ${\tilde W}$ 
and use the recursion relation for it. The contribution of the 
three-$W$ vertex gives
\bea
&& \frac{i g}{\sqrt{2}}  
\left(p_{1}+p_{2}\right)\cdot\left[W_{1},W_{2}\right]=\frac{ig}{\sqrt{2}}\left(\left(p_{1}^{2}-m_{W}^{2}\right)-\left(p_{2}^{2}-m_{W}^{2}\right)\right)W_{1}\cdot W_{2}\nonumber\\
&&-\frac{ig}{\sqrt{2}}\frac{1}{m_{W}}\left(\tilde{G}_{1}\left(p_{1}\cdot W_{2}\right)-\tilde{G}_{2}\left(p_{2}\cdot W_{1}\right)\right) =
\frac{g}{\sqrt{2}}\left(\tilde{W}_{1}\cdot W_{2}-W_{1}\cdot\tilde{W}_{2}\right).
\label{eq3w}
\eea
The four-$W$ vertex evaluates to  
\be
\frac{ig^2}{2}\left(p_{1}+p_{2}+p_{3}\right)\cdot\left\{ W_{1},W_{2},W_{3}\right\} =\frac{ig^2}{2} \left \{  
W_{1}\cdot\left[W_{2},W_{3}\right]-\left[W_{1},W_{2}\right]\cdot W_{3} \right \}.
\ee
If we insert the recursion relation for $\tilde{W}_a$ 
into Eq.(\ref{eq3w}) and combine the resulting expressions 
for three- and four-gluon vertex contributions in 
Eq.(\ref{eq:gauge-inv-eq}), we observe that pure-gauge contributions 
cancel out. The remaining contributions to $p \cdot {\tilde W}$ 
necessarily contain the Higgs boson current.  To investigate those terms, 
we write 
\be
p \cdot {\tilde W} = A_1 + A_2,
\ee
where $A_1$ is the sum of 
Higgs-dependent terms shown as ellipses in Eq.(\ref{eq_rr}) and 
$A_2$  is the sum of Higgs-dependent terms that arise when 
the recursion relation for $\tilde{W}_{1,2}$ is inserted  
into Eq.(\ref{eq3w}). Those terms read 
\bea
\label{eq_a1}
A_1 & =& 
\frac{-ig}{2\sqrt{2}}\left(\left(p_{1}^{2}-m_{H}^{2}\right)-\left(p_{2}^{2}-m_{H}^{2}\right)\right)H_{1}H_{2}\nonumber\\
 && +  
\frac{ig}{\sqrt{2}}m_{W}\left(H_{1}\left(p_{1}\cdot W_{2}\right)+H_{2}\left(p_{2}\cdot W_{1}\right)\right)
+ \frac{ig}{\sqrt{2}}\left(\tilde{G}_{2}H_{1}
+\tilde{G}_{1}H_{2}\right)
\\
 && + 
\frac{ig^{2}}{4m_W}\left(\tilde{G}_{1}H_{2}H_{3}+H_{1}H_{2}\tilde{G}_{3}\right)
+ \frac{ig^{2}}{4}\left(
W_{1} \cdot  \left(p_{2} + p_3 \right ) H_{2}H_{3}
+H_{1}H_{2}\left(p_{1} + p_{2} \right ) \cdot W_3
\right);
\nonumber \\
A_2 & =&  \frac{ig^{2}}{4}\left(-H_{2}H_{3}
\left(p_{2} - p_{3} \right ) \cdot W_{1}
+H_{1}H_{2} \left(p_{1} -p_{2} \right ) \cdot W_{3} \right)\nonumber\\
 && +  \frac{ig^{2}}{4}m_{W}\left(-H_{3}\left(W_{1}\cdot W_{2}\right)+H_{1}\left(W_{2}\cdot W_{3}\right)\right)\nonumber\\
 && +  \frac{ig^{3}}{4\sqrt{2}}\left(-\left(W_{1}\cdot W_{2}\right)H_{3}H_{4}+H_{1}H_{2}\left(W_{3}\cdot W_{4}\right)\right).
\eea

Note that the first term in Eq.(\ref{eq_a1}) 
is $\frac{g}{2\sqrt{2}}\left(\tilde{H}_1H_2-H_1\tilde{H}_2\right)$. 
To simplify it, we insert 
recursion relation to eliminate $\tilde{H}_a$. The appearance of 
Goldstone currents $\tilde{G}_a$ is the consequence of applying 
Ward identity to gauge currents of lower multiplicity.
Finally we need the recursion relation for the 
Goldstone boson current; it reads 
\begin{align}
i m_W \tilde G =&  
 - \frac{ig}{\sqrt{2}}m_{W}\left(H_{2}\left(p_{2}\cdot W_{1}\right)
+H_{1}\left(p_{1}\cdot W_{2}\right)\right)
\nn\\
& - \frac{ig}{2\sqrt{2}}\left(\tilde{G}_{1}H_{2}+H_{1}\tilde{G}_{2}\right) 
+  \frac{ig^{2}}{4}m_{W}\left(\left(W_{1}\cdot W_{2}\right)H_{3}-H_{1}\left(W_{2}\cdot W_{3}\right)\right).
\end{align}
Similar to the previous discussion, 
terms with $\tilde{G}_{1,2}$ currents arise because of electroweak Ward 
identity.
After collecting all terms, we finally arrive at the following 
equation 
\begin{align}
 p \cdot\tilde{W} -(-im_W)\tilde{G} =& 
 \frac{ig}{2\sqrt{2}}\left(\tilde{G}_{1}H_{2}+H_{1}\tilde{G}_{2}\right)
+\frac{ig^{3}}{8\sqrt{2}}\left(H_{1}H_{2}\left(W_{3}\cdot W_{4}\right)-\left(W_{1}\cdot W_{2}\right)H_{3}H_{4}\right)
\nn\\
& +  \frac{ig^{2}}{8m_{W}}\left(\tilde{G}_{1}H_{2}H_{3}+H_{1}H_{2}\tilde{G}_{3}+2H_{1}\tilde{G}_{2}H_{3}\right)\nonumber\\
& +  \frac{ig^{2}}{4}\left(H_{2}H_{3}\left(p_{2}\cdot W_{1}\right)+H_{1}H_{2}\left(p_{2}\cdot W_{3}\right)\right.
\nn \\ 
& \left. +H_{1}H_{3}\left(p_{1}\cdot W_{2}\right)+H_{1}H_{3}\left(p_{3}\cdot W_{2}\right)\right).
\end{align}
Eliminating  $\tilde{G}_a$ in the first term by inserting 
recursion relation, and applying  
the Ward identity  for lower-multiplicity   currents,
we find that all terms cancel out exactly. This proves the  assertion 
that the color-ordered Feynman rules that we constructed allow us to define 
color-ordered currents that satisfy electroweak Ward identity.

\section{Conclusion}
\label{sect5}

In this paper, we studied gauge boson scattering amplitudes in 
$SU(2)$ gauge theory, spontaneously 
broken by the Higgs mechanism. We constructed color-ordered 
scattering amplitudes that satisfy electroweak Ward identities  
and respect perturbative unitarity. 
Those color-ordered amplitudes are peculiar in that 
both external gauge bosons -- that carry 
color -- and the Higgs boson -- that is neutral -- are physically ordered. 
We present explicitly a set of color-ordered Feynman rules, 
which lead to coupled Berends-Giele 
recursion relations for color-ordered currents. 
Similar to QCD, these color-ordered currents 
can be used to efficiently compute tree color-ordered amplitudes. 
We presented a proof of gauge invariance for off-shell currents of 
arbitrary multiplicity. 
Full color-dressed tree-level scattering amplitudes 
can be constructed from color-ordered 
amplitudes  via the  color decomposition in terms of traces of products 
of group generators in the fundamental representation of $SU(2)$.

Our decomposition is restricted to $SU(2)$, due to the  relative simple group structure. 
For a gauge theory of broken $SU(N)$ with a variety of different breaking schemes, the surviving 
global symmetry can be very different. As the result, 
the usefulness of color decomposition  and color-ordered 
amplitudes in that case can be questioned. In particular, 
in realistic electroweak models,  gauge bosons acquire different masses and have 
different couplings to the Higgs sector, so there is a lack of symmetries to make use of.

 Another issue is the  generalization of our decomposition to  the one-loop level. The central
 question is how to construct gauge-invariant color-stripped objects that 
properly reflect the cut structure of the full amplitude and how to assemble those objects 
into the full one-loop amplitude. The unitarity gauge, which deals with only the physical 
degrees of freedom in intermediate states, is clearly ideal for the on-shell methods.
However, as we emphasized several times, our construction of color-stripped amplitudes 
introduces unphysical fields and unphysical vertices whose contribution 
to the entire amplitude  cancels out  once the sum over colors is taken. It is unclear to us at the moment 
if this can be also arranged at the one-loop level.  This remains an interesting open question for the 
future.

\acknowledgments
K.M. would like to thank Zoltan Kunszt for useful discussions. 
This research is supported by the NSF under grants PHY-0855365 
and  by the start-up funds
provided by the Johns Hopkins University. L.D. is supported 
by the Rowland Research Fellowship awarded  by the Department of Physics 
and Astronomy of the Johns Hopkins University.

\newpage

\appendix

\section{Relevant Feynman rules}
\label{sec:appendix_fr}
We list all non-vanishing color-ordered Feynman rules in the unitary gauge, which give gauge invariant and unitary partial amplitudes. Gauge bosons, the Higgs boson, and Goldstone bosons are represented by wavy lines, dashed lines and dotted lines, respectively. External legs are ordered clockwise and all momenta are outgoing.
\\
\bea
\label{eq:ordered-rule}
\parbox{4cm}{
  \begin{fmffile}{3W}
    \begin{fmfgraph*}(60,60)
      %\fmfpen{thick}
      \fmfleft{v1,v2}
      \fmfright{v3}
      \fmflabel{$\mu$}{v1}
      \fmflabel{$\nu$}{v2}
      \fmflabel{$\rho$}{v3}
      \fmf{boson,label=$k_1$}{v1,i1}
      \fmf{boson,label=$k_2$}{v2,i1}
      \fmf{boson,label=$k_3$}{v3,i1}
      \fmfdot{i1}
    \end{fmfgraph*}
  \end{fmffile}
}
&&\hspace*{-1cm}=\frac{ig}{\sqrt{2}}\left[g^{\mu\nu}\left(k_1-k_2\right)^{\rho}+g^{\nu\rho}\left(k_2-k_3\right)^{\mu}+g^{\rho\mu}\left(k_3-k_1\right)^{\nu}\right] 
\nonumber\\
&&
\nonumber\\
\nonumber\\
\parbox{4cm}{
  \begin{fmffile}{4W}
    \begin{fmfgraph*}(80,60)
      %\fmfpen{thick}
      \fmfleft{v1,v2}
      \fmfright{v4,v3}
      \fmflabel{$\mu$}{v1}
      \fmflabel{$\nu$}{v2}
      \fmflabel{$\rho$}{v3}
      \fmflabel{$\sigma$}{v4}
      \fmf{boson}{v1,i1}
      \fmf{boson}{v2,i1}
      \fmf{boson}{v3,i1}
      \fmf{boson}{v4,i1}
      \fmfdot{i1}
    \end{fmfgraph*}
  \end{fmffile}
}
&&\hspace*{-1cm}=ig^2\left[g^{\mu\rho}g^{\nu\sigma}-\frac{1}{2}g^{\mu\nu}g^{\rho\sigma}-\frac{1}{2}g^{\mu\sigma}g^{\nu\rho}\right]
\nonumber\\
&&
\nonumber\\
\nonumber\\
\parbox{4cm}{
  \begin{fmffile}{2WH}
    \begin{fmfgraph*}(60,60)
      %\fmfpen{thick}
      \fmfleft{v1,v2}
      \fmfright{v3}
      \fmflabel{$\mu$}{v1}
      \fmflabel{$\nu$}{v2}
      \fmf{boson}{v1,i1}
      \fmf{boson}{v2,i1}
      \fmf{dashes}{v3,i1}
      \fmfdot{i1}
    \end{fmfgraph*}
  \end{fmffile}
}
&&\hspace*{-1.5cm}=\frac{ig}{\sqrt{2}}m_{W}g^{\mu\nu}
\qquad \qquad
\parbox{4cm}{
  \begin{fmffile}{W2H}
    \begin{fmfgraph*}(60,60)
      %\fmfpen{thick}
      \fmfleft{v1}
      \fmfright{v3,v2}
      \fmflabel{$\mu$}{v1}
      \fmf{boson}{v1,i1}
      \fmf{dashes,label=$p_1$}{v2,i1}
      \fmf{dashes,label=$p_2$}{v3,i1}
      \fmfdot{i1}
    \end{fmfgraph*}
  \end{fmffile}
}
\hspace*{-1.5cm}=-\frac{ig}{2\sqrt{2}}\left(p_1-p_2\right)^{\mu}
\nonumber\\
&&
\nonumber\\
\nonumber\\
\parbox{4cm}{
  \begin{fmffile}{2WHH1}
    \begin{fmfgraph*}(80,60)
      %\fmfpen{thick}
      \fmfleft{v1,v2}
      \fmfright{v4,v3}
      \fmflabel{$\mu$}{v1}
      \fmflabel{$\nu$}{v2}
      \fmf{boson}{v1,i1}
      \fmf{boson}{v2,i1}
      \fmf{dashes}{v3,i1}
      \fmf{dashes}{v4,i1}
      \fmfdot{i1}
    \end{fmfgraph*}
  \end{fmffile}
}
&&\hspace*{-1cm}=\frac{ig^2}{4}g^{\mu\nu}
\nonumber\\
&&
\nonumber\\
\nonumber\\
\parbox{4cm}{
  \begin{fmffile}{2WGH1}
    \begin{fmfgraph*}(80,60)
      %\fmfpen{thick}
      \fmfleft{v1,v2}
      \fmfright{v4,v3}
      \fmflabel{$\mu$}{v1}
      \fmflabel{$\nu$}{v2}
      \fmf{boson}{v1,i1}
      \fmf{boson}{v2,i1}
      \fmf{dots}{v3,i1}
      \fmf{dashes}{v4,i1}
      \fmfdot{i1}
    \end{fmfgraph*}
  \end{fmffile}
}
&&\hspace*{-1cm}=-\frac{g^2}{4}g^{\mu\nu}
\qquad \qquad
\parbox{4cm}{
  \begin{fmffile}{2WGH2}
    \begin{fmfgraph*}(80,60)
      %\fmfpen{thick}
      \fmfleft{v1,v2}
      \fmfright{v4,v3}
      \fmflabel{$\mu$}{v2}
      \fmflabel{$\nu$}{v3}
      \fmf{boson}{v2,i1}
      \fmf{boson}{v3,i1}
      \fmf{dots}{v1,i1}
      \fmf{dashes}{v4,i1}
      \fmfdot{i1}
    \end{fmfgraph*}
  \end{fmffile}
}
\hspace*{-1cm}=\frac{g^2}{4}g^{\mu\nu}
\nonumber\\
&&
\nonumber\\
\nonumber\\
\parbox{4cm}{
  \begin{fmffile}{WGH1}
    \begin{fmfgraph*}(60,60)
      \fmfleft{v1}
      \fmfright{v3,v2}
      \fmflabel{$\mu$}{v1}
      \fmf{boson}{v1,i1}
      \fmf{dots,label=$p_1$}{v2,i1}
      \fmf{dashes,label=$p_2$}{v3,i1}
      \fmfdot{i1}
    \end{fmfgraph*}
  \end{fmffile}
}
&&\hspace*{-1.5cm}=\frac{g}{2\sqrt{2}}\left(p_1-p_2\right)^{\mu}
\qquad \qquad
\parbox{4cm}{
  \begin{fmffile}{WGH2}
    \begin{fmfgraph*}(60,60)
      \fmfleft{v1}
      \fmfright{v3,v2}
      \fmflabel{$\mu$}{v1}
      \fmf{boson}{v1,i1}
      \fmf{dots,label=$p_1$}{v3,i1}
      \fmf{dashes,label=$p_2$}{v2,i1}
      \fmfdot{i1}
    \end{fmfgraph*}
  \end{fmffile}
}
\hspace*{-1.5cm}=\frac{g}{2\sqrt{2}}\left(p_1-p_2\right)^{\mu}
\nonumber\\
\eea

We also present below a set of color-dressed Feynman rules for the extended particle content, which is useful in the proof of color decomposition.

\bea
\label{eq:dressed-rule}
\parbox{4cm}{
  \begin{fmffile}{3W_cl}
    \begin{fmfgraph*}(60,60)
      \fmfleft{v1,v2}
      \fmfright{v3}
      \fmflabel{$\mu, \tilde{a}$}{v1}
      \fmflabel{$\nu, \tilde{b}$}{v2}
      \fmflabel{$\rho, \tilde{c}$}{v3}
      \fmf{boson,label=$k_1$}{v1,i1}
      \fmf{boson,label=$k_2$}{v2,i1}
      \fmf{boson,label=$k_3$}{v3,i1}
      \fmfdot{i1}
    \end{fmfgraph*}
  \end{fmffile}
}
&&\hspace*{-1cm}=-g\varepsilon^{\tilde{a}\tilde{b}\tilde{c}}\left[g^{\mu\nu}\left(k_1-k_2\right)^{\rho}+g^{\nu\rho}\left(k_2-k_3\right)^{\mu}+g^{\rho\mu}\left(k_3-k_1\right)^{\nu}\right] 
\nonumber\\
&&
\nonumber\\
\nonumber\\
\nonumber\\
\parbox{4cm}{
  \begin{fmffile}{4W_cl}
    \begin{fmfgraph*}(80,60)
      %\fmfpen{thick}
      \fmfleft{v1,v2}
      \fmfright{v4,v3}
      \fmflabel{$\mu,\tilde a$}{v1}
      \fmflabel{$\nu,\tilde b$}{v2}
      \fmflabel{$\rho,\tilde c$}{v3}
      \fmflabel{$\sigma,\tilde d$}{v4}
      \fmf{boson}{v1,i1}
      \fmf{boson}{v2,i1}
      \fmf{boson}{v3,i1}
      \fmf{boson}{v4,i1}
      \fmfdot{i1}
    \end{fmfgraph*}
  \end{fmffile}
}
&&\hspace*{-1cm}\begin{array}{c}=-ig^2\left[\varepsilon^{\tilde a \tilde b \tilde e}\varepsilon^{\tilde e \tilde c \tilde d}\left(g^{\mu\rho}g^{\nu\sigma}-g^{\mu\sigma}g^{\nu\rho}\right)\right.\\
\left.+\varepsilon^{\tilde a \tilde c \tilde e}\varepsilon^{ \tilde e \tilde b \tilde d}\left(g^{\mu\nu}g^{\rho\sigma}-g^{\mu\sigma}g^{\nu\rho}\right)\right.\\
\left.+\varepsilon^{ \tilde a \tilde d \tilde e}\varepsilon^{ \tilde e \tilde b \tilde c}\left(g^{\mu\nu}g^{\rho\sigma}-g^{\mu\rho}g^{\nu\sigma}\right)\right]\end{array}
\nonumber\\
&&
\nonumber\\
\nonumber\\
\nonumber\\
\parbox{4cm}{
  \begin{fmffile}{WWH_cl}
    \vskip0.1cm
    \begin{fmfgraph*}(60,60)
      \fmfleft{v1,v2}
      \fmfright{v3}
      \fmflabel{$\mu, \tilde a$}{v1}
      \fmflabel{$\nu, \tilde b$}{v2}
      \fmflabel{$\tilde c$}{v3}
      \fmf{boson}{v1,i1}
      \fmf{boson}{v2,i1}
      \fmf{dashes}{v3,i1}
      \fmfdot{i1}
    \end{fmfgraph*}
  \end{fmffile}
}
&&\hspace*{-1cm}=\frac{ig}{\sqrt{2}}m_Wg^{\mu\nu}\left\{\mathrm{Tr}\left(T^{\tilde a}T^{\tilde b}T^{\tilde c}\right)+\mathrm{Tr}\left(T^{\tilde b}T^{\tilde a}T^{\tilde c}\right)\right\}
\nonumber\\
&&
\nonumber\\
\nonumber\\
\nonumber\\
\parbox{4cm}{
  \begin{fmffile}{WWHH_cl}
    \vskip0.15cm
    \begin{fmfgraph*}(80,60)
      \fmfleft{v1,v2}
      \fmfright{v4,v3}
      \fmflabel{$\mu,\tilde a$}{v1}
      \fmflabel{$\nu,\tilde b$}{v2}
      \fmflabel{$\tilde c$}{v3}
      \fmflabel{$\tilde d$}{v4}
      \fmf{boson}{v1,i1}
      \fmf{boson}{v2,i1}
      \fmf{dashes}{v3,i1}
      \fmf{dashes}{v4,i1}
      \fmfdot{i1}
    \end{fmfgraph*}
  \end{fmffile}
}
&&\hspace*{-1cm}\begin{array}{cc}=\frac{ig^{2}}{4}g^{\mu\nu}\left\{\left(\mathrm{Tr}\left(T^{\tilde a}T^{\tilde b}T^{\tilde e}\right)+\mathrm{Tr}\left(T^{\tilde b}T^{\tilde a}T^{\tilde e}\right)\right)\right.\\
\left.\times\left(\mathrm{Tr}\left(T^{\tilde c}T^{\tilde d}T^{\tilde e}\right)+\mathrm{Tr}\left(T^{\tilde d}T^{\tilde c}T^{\tilde e}\right)\right)\right\}\end{array}
\nonumber\\
&&
\nonumber\\
\nonumber\\
\parbox{4cm}{
  \begin{fmffile}{WGH_cl}
    \begin{fmfgraph*}(60,60)
      \fmfleft{v1}
      \fmfright{v3,v2}
      \fmflabel{$\mu,\tilde a$}{v1}
      \fmflabel{$\tilde b$}{v2}
      \fmflabel{$\tilde c$}{v3}
      \fmf{boson}{v1,i1}
      \fmf{dots,label=$p_1$}{v2,i1}
      \fmf{dashes,label=$p_2$}{v3,i1}
      \fmfdot{i1}
    \end{fmfgraph*}
  \end{fmffile}
}
&&\hspace*{-1cm}=-\frac{g}{2\sqrt{2}}\left(p_1-p_2\right)^{\mu}\left\{\mathrm{Tr}\left(T^{\tilde a}T^{\tilde b}T^{\tilde c}\right)+\mathrm{Tr}\left(T^{\tilde b}T^{\tilde a}T^{\tilde c}\right)\right\}
\nonumber\\
&&
\nonumber\\
\nonumber\\
\nonumber\\
\parbox{4cm}{
  \begin{fmffile}{WHH_cl}
    \vskip0.15cm
    \begin{fmfgraph*}(60,60)
      \fmfleft{v1}
      \fmfright{v3,v2}
      \fmflabel{$\mu,\tilde a$}{v1}
      \fmflabel{$\tilde b$}{v2}
      \fmflabel{$\tilde c$}{v3}
      \fmf{boson}{v1,i1}
      \fmf{dashes,label=$p_1$}{v2,i1}
      \fmf{dashes,label=$p_2$}{v3,i1}
      \fmfdot{i1}
    \end{fmfgraph*}
  \end{fmffile}
}
&&\hspace*{-1cm}=-i\frac{g}{2\sqrt{2}}\left(p_1-p_2\right)^{\mu}\left\{\mathrm{Tr}\left(T^{\tilde a}T^{\tilde b}T^{\tilde c}\right)-\mathrm{Tr}\left(T^{\tilde b}T^{\tilde a}T^{\tilde c}\right)\right\}
\nonumber\\
&&
\nonumber\\
\nonumber\\
\parbox{4cm}{
  \begin{fmffile}{WWGH_cl}
    \begin{fmfgraph*}(80,60)
      %\fmfpen{thick}
      \fmfleft{v1,v2}
      \fmfright{v4,v3}
      \fmflabel{$\mu,\tilde a$}{v1}
      \fmflabel{$\nu,\tilde b$}{v2}
      \fmflabel{{$\tilde c$}}{v3}
      \fmflabel{$\tilde d$}{v4}
      \fmf{boson}{v1,i1}
      \fmf{boson}{v2,i1}
      \fmf{dots}{v3,i1}
      \fmf{dashes}{v4,i1}
      \fmfdot{i1}
    \end{fmfgraph*}
  \end{fmffile}
}
&&\hspace*{-1cm}\begin{array}{cc}=-\frac{g^2}{4}g^{\mu\nu}
\left\{\left(\mathrm{Tr}\left(T^{\tilde a}T^{ \tilde b}T^{\tilde e}\right)+\mathrm{Tr}\left(T^{\tilde b}T^{\tilde a}T^{\tilde e}\right)\right)\right.\\
\left.\times\left(\mathrm{Tr}\left(T^{\tilde e}T^{ \tilde c}T^{\tilde d}\right)-\mathrm{Tr}\left(T^{\tilde c}T^{\tilde e}T^{\tilde d}\right)\right)
\right\}\end{array}
\nonumber\\
\eea

\section{The Ward identity}
\label{sec:appendix_wy}

\begin{figure}[h!]
\centering
\includegraphics[scale=0.6]{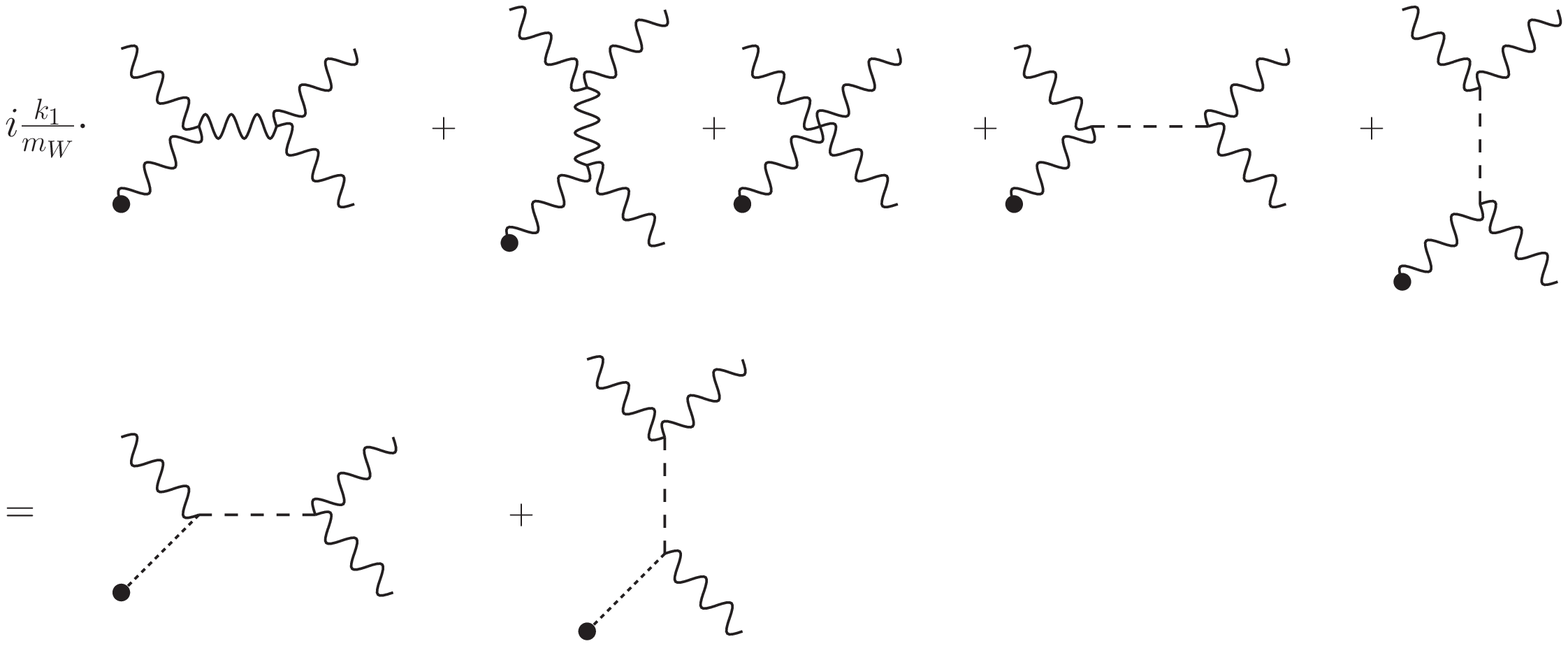}
\caption{\label{fig:WWWW} Gauge invariance for $W W \rightarrow W W $ color-ordered  amplitude.}
\end{figure}

\begin{figure}[h!]
\centering
\includegraphics[scale=0.6]{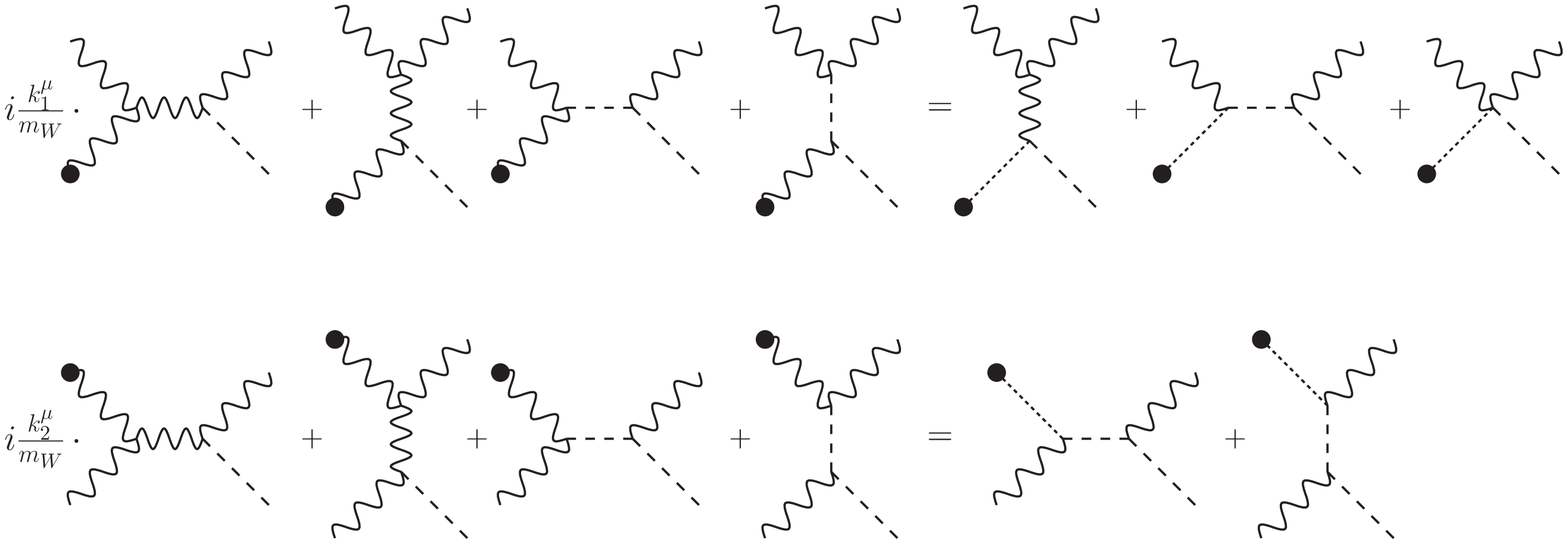}
\caption{\label{fig:WWWH} Gauge invariance for $W W\rightarrow WH$ partial amplitude.}
\end{figure}

\begin{figure}[h!]
\centering
\includegraphics[scale=0.6]{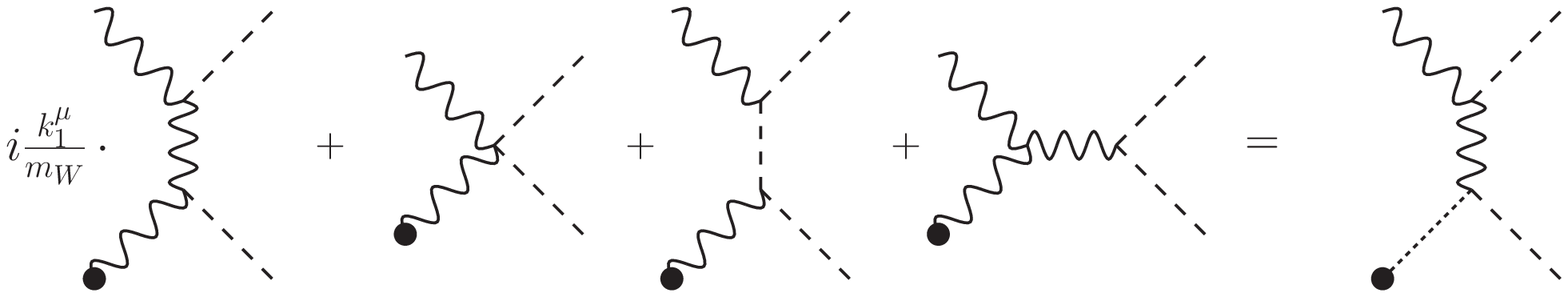}
\caption{\label{fig:WWHH} Gauge invariance for $W W\rightarrow HH$ partial amplitude.}
\end{figure}

\begin{figure}[h!]
\centering
\includegraphics[scale=0.6]{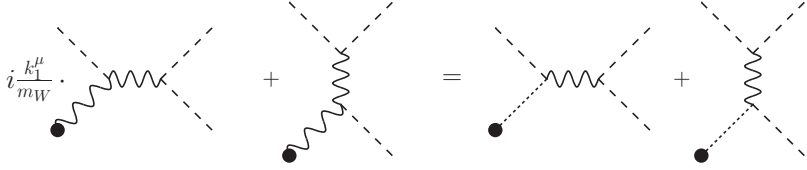}
\caption{\label{fig:WHHH} Gauge invariance for $WH\rightarrow HH$ partial amplitude.}
\end{figure}

\newpage

\section{Numerical results for amplitudes}

For future reference, we present numerical results for multi-$W$ scattering amplitudes for a  
typical phase space point. We choose $m_W = 80~{\rm GeV}$ and 
$m_H = 114~{\rm GeV}$ and set the gauge coupling constant to $g = 0.1$. All momenta are given in $\rm GeV$.

\subsection*{Scattering of five $W$ bosons}
We first list all color-ordered primitive amplitudes (there are $4!=24$ of them) for $W(k_1)W(k_2)\rightarrow W(k_3)W(k_4)W(k_5)$ scattering. 
In the center of mass frame, the momenta of gauge bosons are 
\bea
k_1^{\mu}&=&\left(\right.\mathtt{400.0},\,\mathtt{0.0},\left.\mathtt{0.0},\,\mathtt{391.91835884531}\right)\nonumber\\
k_2^{\mu}&=&\left(\right.\mathtt{400.0},\,\mathtt{0.0},\left.\mathtt{0.0},\,\mathtt{-391.91835884531}\right)\nonumber\\
k_3^{\mu}&=&\left(\right.\mathtt{141.60000091791},\,\mathtt{-21.221509298529},
\left.\mathtt{-65.313093963515},\,\mathtt{-94.521995112027}\right)\nonumber\\
k_4^{\mu}&=&\left(\right.\mathtt{272.59123471245},\,\mathtt{-47.332020563805},
\left.\mathtt{-145.67298975201},\,\mathtt{-210.81992583272}\right)\nonumber\\
k_5^{\mu}&=&\left(\right.\mathtt{385.80876436964},\,\mathtt{68.553529862334},
\left.\mathtt{210.98608371552},\,\mathtt{305.34192094475}\right)\nonumber\\
\eea
The first benchmark configuration has all longitudinal polarizations $\left\{h_i\right\}=\left\{L,L,L,L,L\right\}$\footnote{To avoid confusion, throughout this paper we do not take complex conjugate when multiplying polarization vectors, regardless of whether the external particle is incoming or outgoing.}
\bea
\epsilon_1^{\mu}&=&\left(\right.\mathtt{4.8989794855664},\,\mathtt{0.0},\left.\mathtt{0.0},\,\mathtt{5.0}\right)\nonumber\\
\epsilon_2^{\mu}&=&\left(\right.\mathtt{4.8989794855664},\,\mathtt{0.0},\left.\mathtt{0.0},\,\mathtt{-5.0}\right)\nonumber\\
\epsilon_3^{\mu}&=&\left(\right.\mathtt{1.4604451515266},\,\mathtt{-0.32149505634144},
\left.\mathtt{-0.98946010522866},\,\mathtt{-1.4319600795855}\right)\nonumber\\
\epsilon_4^{\mu}&=&\left(\right.\mathtt{3.2573470139166},\,\mathtt{-0.61890348724549},
\left.\mathtt{-1.9047891951593},\,\mathtt{-2.7566367487485}\right)\nonumber\\
\epsilon_5^{\mu}&=&\left(\right.\mathtt{4.7177921654432},\,\mathtt{0.87595769515527},
\left.\mathtt{2.6959207494118},\,\mathtt{3.9015730603830}\right)\nonumber\\
\eea 
Color-ordered primitive amplitudes are evaluated to be
\bea
A^{\mathrm{tree}}_5\left(1,2,3,4,5\right)&=&\mathtt{8.082744225626406}\,\times 10^{-6}~\mathrm{GeV^{-1}}\nonumber\\
A^{\mathrm{tree}}_5\left(1,2,3,5,4\right)&=&\mathtt{-9.052099561980900}\,\times 10^{-7}~\mathrm{GeV^{-1}}\nonumber\\
A^{\mathrm{tree}}_5\left(1,2,4,3,5\right)&=&\mathtt{2.485269458065873}\,\times 10^{-5}~\mathrm{GeV^{-1}}\nonumber\\
A^{\mathrm{tree}}_5\left(1,2,4,5,3\right)&=&\mathtt{3.859247304761025}\,\times 10^{-7}~\mathrm{GeV^{-1}}\nonumber\\
A^{\mathrm{tree}}_5\left(1,2,5,3,4\right)&=&\mathtt{6.760721099941013}\,\times 10^{-7}~\mathrm{GeV^{-1}}\nonumber\\
A^{\mathrm{tree}}_5\left(1,2,5,4,3\right)&=&\mathtt{-1.552100997888805}\,\times 10^{-6}~\mathrm{GeV^{-1}}\nonumber\\
A^{\mathrm{tree}}_5\left(1,3,2,4,5\right)&=&\mathtt{-7.534154547657562}\,\times 10^{-6}~\mathrm{GeV^{-1}}\nonumber\\
A^{\mathrm{tree}}_5\left(1,3,2,5,4\right)&=&\mathtt{3.909646009037787}\,\times 10^{-7}~\mathrm{GeV^{-1}}\nonumber\\
A^{\mathrm{tree}}_5\left(1,3,4,2,5\right)&=&\mathtt{2.741897891315819}\,\times 10^{-5}~\mathrm{GeV^{-1}}\nonumber\\
A^{\mathrm{tree}}_5\left(1,3,4,5,2\right)&=&\mathtt{1.552100997890106}\,\times 10^{-6}~\mathrm{GeV^{-1}}\nonumber\\
A^{\mathrm{tree}}_5\left(1,3,5,2,4\right)&=&\mathtt{7.476469442186263}\,\times 10^{-7}~\mathrm{GeV^{-1}}\nonumber\\
A^{\mathrm{tree}}_5\left(1,3,5,4,2\right)&=&\mathtt{-3.859247304756688}\,\times 10^{-7}~\mathrm{GeV^{-1}}\nonumber\\
A^{\mathrm{tree}}_5\left(1,4,2,3,5\right)&=&\mathtt{-1.052143197490972}\,\times 10^{-5}~\mathrm{GeV^{-1}}\nonumber\\
A^{\mathrm{tree}}_5\left(1,4,2,5,3\right)&=&\mathtt{-7.476469442214452}\,\times 10^{-7}~\mathrm{GeV^{-1}}\nonumber\\
A^{\mathrm{tree}}_5\left(1,4,3,2,5\right)&=&\mathtt{6.344716114626181}\,\times 10^{-6}~\mathrm{GeV^{-1}}\nonumber\\
A^{\mathrm{tree}}_5\left(1,4,3,5,2\right)&=&\mathtt{-6.760721099878129}\,\times 10^{-7}~\mathrm{GeV^{-1}}\nonumber\\
A^{\mathrm{tree}}_5\left(1,4,5,2,3\right)&=&\mathtt{-3.909646009018272}\,\times 10^{-7}~\mathrm{GeV^{-1}}\nonumber\\
A^{\mathrm{tree}}_5\left(1,4,5,3,2\right)&=&\mathtt{9.052099561980900}\,\times 10^{-7}~\mathrm{GeV^{-1}}\nonumber\\
A^{\mathrm{tree}}_5\left(1,5,2,3,4\right)&=&\mathtt{-6.344716114630626}\,\times 10^{-6}~\mathrm{GeV^{-1}}\nonumber\\
A^{\mathrm{tree}}_5\left(1,5,2,4,3\right)&=&\mathtt{-2.741897891315862}\,\times 10^{-5}~\mathrm{GeV^{-1}}\nonumber\\
A^{\mathrm{tree}}_5\left(1,5,3,2,4\right)&=&\mathtt{1.052143197490939}\,\times 10^{-5}~\mathrm{GeV^{-1}}\nonumber\\
A^{\mathrm{tree}}_5\left(1,5,3,4,2\right)&=&\mathtt{-2.485269458066134}\,\times 10^{-5}~\mathrm{GeV^{-1}}\nonumber\\
A^{\mathrm{tree}}_5\left(1,5,4,2,3\right)&=&\mathtt{7.534154547657562}\,\times 10^{-6}~\mathrm{GeV^{-1}}\nonumber\\
A^{\mathrm{tree}}_5\left(1,5,4,3,2\right)&=&\mathtt{-8.082744225625105}\,\times 10^{-6}~\mathrm{GeV^{-1}}\nonumber\\
\eea
The second benchmark configuration has both transverse and longitudinal polarizations $\left\{h_i\right\}=\left\{+,-,+,L,L\right\}$. The polarization vectors are
\bea
\epsilon_1^{\mu}&=&\left(\right.\mathtt{0.0},\,\mathtt{-0.70710678118655},\left.\mathtt{0.70710678118655}i,\,\mathtt{0.0}\right)\nonumber\\
\epsilon_2^{\mu}&=&\left(\right.\mathtt{0.0},\,\mathtt{0.70710678118655},\left.\mathtt{-0.70710678118655}i,\,\mathtt{0.0}\right)\nonumber\\
\epsilon_3^{\mu}&=&\left(\right.\mathtt{0.0},\,\mathtt{0.17677668390637-0.67249851605560}i,\nonumber\\
&&\left.\mathtt{0.54406272447999+0.21850799963164}i,\,\mathtt{-0.41562694313348}\right)\nonumber\\
\epsilon_4^{\mu}&=&\left(\right.\mathtt{3.2573470139166},\,\mathtt{-0.61890348724549},
\left.\mathtt{-1.9047891951593},\,\mathtt{-2.7566367487485}\right)\nonumber\\
\epsilon_5^{\mu}&=&\left(\right.\mathtt{4.7177921654432},\,\mathtt{0.87595769515527},
\left.\mathtt{2.6959207494118},\,\mathtt{3.9015730603830}\right)\nonumber\\
\eea
Color-ordered primitive amplitudes are evaluated to be
\bea
A^{\mathrm{tree}}_5\left(1,2,3,4,5\right)&=&(\mathtt{-1.587502368520595E-1.153387893203229}i)\,\times 10^{-6}~\mathrm{GeV^{-1}}\nonumber\\
A^{\mathrm{tree}}_5\left(1,2,3,5,4\right)&=&(\mathtt{-5.687601324129297-4.132283918887344}i)\,\times 10^{-8}~\mathrm{GeV^{-1}}\nonumber\\
A^{\mathrm{tree}}_5\left(1,2,4,3,5\right)&=&(\mathtt{-1.715712547908148-1.246538033715343}i)\,\times 10^{-6}~\mathrm{GeV^{-1}}\nonumber\\
A^{\mathrm{tree}}_5\left(1,2,4,5,3\right)&=&(\mathtt{-1.881276398229126-1.366827202600273}i)\,\times 10^{-7}~\mathrm{GeV^{-1}}\nonumber\\
A^{\mathrm{tree}}_5\left(1,2,5,3,4\right)&=&(\mathtt{2.773037381457798+2.014729430705663}i)\,\times 10^{-7}~\mathrm{GeV^{-1}}\nonumber\\
A^{\mathrm{tree}}_5\left(1,2,5,4,3\right)&=&(\mathtt{4.716241589780851+3.426549817461441}i)\,\times 10^{-8}~\mathrm{GeV^{-1}}\nonumber\\
A^{\mathrm{tree}}_5\left(1,3,2,4,5\right)&=&(\mathtt{2.433008233863460+1.767683813673294}i)\,\times 10^{-6}~\mathrm{GeV^{-1}}\nonumber\\
A^{\mathrm{tree}}_5\left(1,3,2,5,4\right)&=&(\mathtt{1.085894764699578+0.7889486653585891}i)\,\times 10^{-7}~\mathrm{GeV^{-1}}\nonumber\\
A^{\mathrm{tree}}_5\left(1,3,4,2,5\right)&=&(\mathtt{-1.188112309463586-0.8632140528886585}i)\,\times 10^{-6}~\mathrm{GeV^{-1}}\nonumber\\
A^{\mathrm{tree}}_5\left(1,3,4,5,2\right)&=&(\mathtt{-4.716241589779813-3.426549817461366}i)\,\times 10^{-8}~\mathrm{GeV^{-1}}\nonumber\\
A^{\mathrm{tree}}_5\left(1,3,5,2,4\right)&=&(\mathtt{2.529171126417367+1.837550239227348}i)\,\times 10^{-7}~\mathrm{GeV^{-1}}\nonumber\\
A^{\mathrm{tree}}_5\left(1,3,5,4,2\right)&=&(\mathtt{1.881276398229151+1.366827202600311}i)\,\times 10^{-7}~\mathrm{GeV^{-1}}\nonumber\\
A^{\mathrm{tree}}_5\left(1,4,2,3,5\right)&=&(\mathtt{-2.800106605902566-2.034396371916651}i)\,\times 10^{-7}~\mathrm{GeV^{-1}}\nonumber\\
A^{\mathrm{tree}}_5\left(1,4,2,5,3\right)&=&(\mathtt{-2.529171126417509-1.837550239227342}i)\,\times 10^{-7}~\mathrm{GeV^{-1}}\nonumber\\
A^{\mathrm{tree}}_5\left(1,4,3,2,5\right)&=&(\mathtt{1.312085088984693+0.9532855424320996}i)\,\times 10^{-6}~\mathrm{GeV^{-1}}\nonumber\\
A^{\mathrm{tree}}_5\left(1,4,3,5,2\right)&=&(\mathtt{-2.773037381457778-2.014729430705651}i)\,\times 10^{-7}~\mathrm{GeV^{-1}}\nonumber\\
A^{\mathrm{tree}}_5\left(1,4,5,2,3\right)&=&(\mathtt{-1.085894764699594-0.7889486653588198}i)\,\times 10^{-7}~\mathrm{GeV^{-1}}\nonumber\\
A^{\mathrm{tree}}_5\left(1,4,5,3,2\right)&=&(\mathtt{5.687601324147984+4.132283918887015}i)\,\times 10^{-8}~\mathrm{GeV^{-1}}\nonumber\\
A^{\mathrm{tree}}_5\left(1,5,2,3,4\right)&=&(\mathtt{-1.312085088984696-0.9532855424320850}i)\,\times 10^{-6}~\mathrm{GeV^{-1}}\nonumber\\
A^{\mathrm{tree}}_5\left(1,5,2,4,3\right)&=&(\mathtt{1.188112309463583+0.8632140528886511}i)\,\times 10^{-6}~\mathrm{GeV^{-1}}\nonumber\\
A^{\mathrm{tree}}_5\left(1,5,3,2,4\right)&=&(\mathtt{2.800106605902975+2.034396371916417}i)\,\times 10^{-7}~\mathrm{GeV^{-1}}\nonumber\\
A^{\mathrm{tree}}_5\left(1,5,3,4,2\right)&=&(\mathtt{1.715712547908162+1.246538033715348}i)\,\times 10^{-6}~\mathrm{GeV^{-1}}\nonumber\\
A^{\mathrm{tree}}_5\left(1,5,4,2,3\right)&=&(\mathtt{-2.433008233863598-1.767683813673305}i)\,\times 10^{-6}~\mathrm{GeV^{-1}}\nonumber\\
A^{\mathrm{tree}}_5\left(1,5,4,3,2\right)&=&(\mathtt{1.587502368520560+1.153387893203234}i)\,\times 10^{-6}~\mathrm{GeV^{-1}}\nonumber\\
\eea

We point out that the results exhibit reflection symmetry of the primitive amplitudes for general $W$-Higgs multi-particle scattering
\be
A_{n}^{\mathrm{tree}}\left(1,2,\cdots,n-1,n\right)=\left(-1\right)^{m}A_{n}^{\mathrm{tree}}\left(n,n-1,\cdots,2,1\right)
\ee
where $m$ is the number of external gauge bosons $W$.

\subsection*{Scattering of six $W$ bosons}
Next we present numerical results for full amplitude of 6-$W$ scattering, i.e. $W(k_1)W(k_2)\rightarrow W(k_3)W(k_4)W(k_5)W(k_6)$. Their momenta in center of mass frame are 
\bea
k_1^{\mu}&=&\left(\right.\mathtt{400.0},\,\mathtt{0.0},\left.\mathtt{0.0},\,\mathtt{391.91835884531}\right)\nonumber\\
k_2^{\mu}&=&\left(\right.\mathtt{400.0},\,\mathtt{0.0},\left.\mathtt{0.0},\,\mathtt{-391.91835884531}\right)\nonumber\\
k_3^{\mu}&=&\left(\right.\mathtt{137.60000085831},\,\mathtt{-20.334885027467},
\left.\mathtt{-62.584344866943},\,\mathtt{-90.572912422541}\right)\nonumber\\
k_4^{\mu}&=&\left(\right.\mathtt{111.42091925759},\,\mathtt{-14.086567220910},
\left.\mathtt{-43.353998793403},\,\mathtt{-62.742494856025}\right)\nonumber\\
k_5^{\mu}&=&\left(\right.\mathtt{185.38050614865},\,\mathtt{-30.374957508527},
\left.\mathtt{-93.484512622753},\,\mathtt{-135.29205414942}\right)\nonumber\\
k_6^{\mu}&=&\left(\right.\mathtt{365.59857373546},\,\mathtt{64.796409756903},
\left.\mathtt{199.42285628310},\,\mathtt{288.60746142799}\right)\nonumber\\
\eea
The first benchmark configuration has all longitudinal polarizations $\left\{h_i\right\}=\left\{L,L,L,L,L,L\right\}$
\bea
\epsilon_1^{\mu}&=&\left(\right.\mathtt{4.8989794855664},\,\mathtt{0.0},\left.\mathtt{0.0},\,\mathtt{5.0}\right)\nonumber\\
\epsilon_2^{\mu}&=&\left(\right.\mathtt{4.8989794855664},\,\mathtt{0.0},\left.\mathtt{0.0},\,\mathtt{-5.0}\right)\nonumber\\
\epsilon_3^{\mu}&=&\left(\right.\mathtt{1.3994284679494},\,\mathtt{-0.31241327501240},
\left.\mathtt{-0.96150925456316},\,\mathtt{-1.3915092295391}\right)\nonumber\\
\epsilon_4^{\mu}&=&\left(\right.\mathtt{0.96942486559412},\,\mathtt{-0.25297510227489},
\left.\mathtt{-0.77857735719366},\,\mathtt{-1.1267677074386}\right)\nonumber\\
\epsilon_5^{\mu}&=&\left(\right.\mathtt{2.0903772110229},\,\mathtt{-0.42089629860535},
\left.\mathtt{-1.2953856916111},\,\mathtt{-1.8746997359986}\right)\nonumber\\
\epsilon_6^{\mu}&=&\left(\right.\mathtt{4.4592305445664},\,\mathtt{0.83007156284956},
\left.\mathtt{3.6971932157406},\,\mathtt{2.5546977464318}\right)\nonumber\\
\eea
To evaluate full amplitude we also have to specify the color of the gauge bosons $\left\{a_i,i=1,2,\cdots,6\right\}$. 
We choose $\left\{a_i\right\}=\left\{1,1,2,2,3,3\right\}$. Then the full color-dressed amplitude is found to be
\be
\mathcal{A}_6^{\mathrm{tree}}\left(1^1_L,2^1_L,3^2_L,4^2_L,5^3_L,6^3_L\right)=\mathtt{1.745869319633557}\,\times 10^{-8}~\mathrm{GeV^{-2}}
\ee
The second benchmark configuration has both transverse and longitudinal polarizations $\left\{h_i\right\}=\left\{+,-,+,L,L,L\right\}$
\bea
\epsilon_1^{\mu}&=&\left(\right.\mathtt{0.0},\,\mathtt{-0.70710678118655},\left.\mathtt{0.70710678118655}i,\,\mathtt{0.0}\right)\nonumber\\
\epsilon_2^{\mu}&=&\left(\right.\mathtt{0.0},\,\mathtt{0.70710678118655},\left.\mathtt{-0.70710678118655}i,\,\mathtt{0.0}\right)\nonumber\\
\epsilon_3^{\mu}&=&\left(\right.\mathtt{0.0},\,\mathtt{0.17677668390637-0.67249851605560}i,\nonumber\\
&&\left.\mathtt{0.54406272447999+0.21850799963164}i,\,\mathtt{-0.41562694313348}\right)\nonumber\\
\epsilon_4^{\mu}&=&\left(\right.\mathtt{0.96942486559412},\,\mathtt{-0.25297510227489},
\left.\mathtt{-0.77857735719366},\,\mathtt{-1.1267677074386}\right)\nonumber\\
\epsilon_5^{\mu}&=&\left(\right.\mathtt{2.0903772110229},\,\mathtt{-0.42089629860535},
\left.\mathtt{-1.2953856916111},\,\mathtt{-1.8746997359986}\right)\nonumber\\
\epsilon_6^{\mu}&=&\left(\right.\mathtt{4.4592305445664},\,\mathtt{0.83007156284956},
\left.\mathtt{3.6971932157406},\,\mathtt{2.5546977464318}\right)\nonumber\\
\eea
Still choosing color $\left\{a_i\right\}=\left\{1,1,2,2,3,3\right\}$, the full color-dressed amplitude evaluates to
\be
\mathcal{A}_6^{\mathrm{tree}}\left(1^1_+,2^1_-,3^2_+,4^2_L,5^3_L,6^3_L\right)=(\mathtt{-5.508936118528462-4.002476058956751}i)\,\times 10^{-9}~\mathrm{GeV^{-2}}
\ee
These results are cross-checked using the conventional Feynman diagrammatic method. In that case we calculate and sum over all 730 tree diagrams for 6-$W$ scattering.

\subsection*{Scattering of nine $W$ bosons}
Finally, we present numerical results for full amplitude of 9-$W$ scattering, i.e. $W(k_1)W(k_2)\rightarrow W(k_3)W(k_4)W(k_5)W(k_6)W(k_7)W(k_8)W(k_9)$. Their momenta in center of mass frame are 
\bea
k_1^{\mu}&=&\left(\right.\mathtt{400.0},\,\mathtt{0.0},\,\mathtt{0.0},\left.\mathtt{391.91835884531}\right)\nonumber\\
k_2^{\mu}&=&\left(\right.\mathtt{400.0},\,\mathtt{0.0},\,\mathtt{0.0},\left.\mathtt{-391.91835884531}\right)\nonumber\\
k_3^{\mu}&=&\left(\right.\mathtt{116.00000053644},\,\mathtt{-15.257392539990},
\left.\mathtt{-46.957428832438},\,\mathtt{-67.957427664543}\right)\nonumber\\
k_4^{\mu}&=&\left(\right.\mathtt{100.88056163819},\,\mathtt{-11.162669509069},
\left.\mathtt{-34.355166367926},\,\mathtt{-49.719262561900}\right)\nonumber\\
k_5^{\mu}&=&\left(\right.\mathtt{89.943319606884},\,\mathtt{-7.4665293771818},
\left.\mathtt{-22.979616008133},\,\mathtt{-33.256411849213}\right)\nonumber\\
k_6^{\mu}&=&\left(\right.\mathtt{82.900106361594},\,\mathtt{-3.9479230294227},
\left.\mathtt{-12.150458487855},\,\mathtt{-17.584308261977}\right)\nonumber\\
k_7^{\mu}&=&\left(\right.\mathtt{80.019253013145},\,\mathtt{-0.31881335310116},
\left.\mathtt{-0.98120667078859},\,\mathtt{-1.4200155973621}\right)\nonumber\\
k_8^{\mu}&=&\left(\right.\mathtt{83.380380566575},\,\mathtt{-4.2685834850712},
\left.\mathtt{-13.137349956107},\,\mathtt{-19.012550975304}\right)\nonumber\\
k_9^{\mu}&=&\left(\right.\mathtt{246.87637827718},\,\mathtt{42.421911293836},
\left.\mathtt{130.56122632325},\,\mathtt{188.94997691030}\right)
\eea
The first benchmark configuration has all longitudinal polarizations $\left\{h_i\right\}=\left\{L,L,L,L,L,L,L,L,L\right\}$
\bea
\epsilon_1^{\mu}&=&\left(\right.\mathtt{4.8989794855664},\,\mathtt{0.0},
\left.\mathtt{0.0},\,\mathtt{5.0000000000000}\right)\nonumber\\
\epsilon_2^{\mu}&=&\left(\right.\mathtt{4.8989794855664},\,\mathtt{0.0},
\left.\mathtt{0.0},\,\mathtt{-5.0000000000000}\right)\nonumber\\
\epsilon_3^{\mu}&=&\left(\right.\mathtt{1.0500000092600},\,\mathtt{-0.26337165583558},
\left.\mathtt{-0.81057466096947},\,\mathtt{-1.1730746392889}\right)\nonumber\\
\epsilon_4^{\mu}&=&\left(\right.\mathtt{0.76820485919648},\,\mathtt{-0.22904379687417},
\left.\mathtt{-0.70492436784595},\,\mathtt{-1.0201761026526}\right)\nonumber\\
\epsilon_5^{\mu}&=&\left(\right.\mathtt{0.51383982516229},\,\mathtt{-0.20421138712643},
\left.\mathtt{-0.62849806430746},\,\mathtt{-0.90957091996858}\right)\nonumber\\
\epsilon_6^{\mu}&=&\left(\right.\mathtt{0.27169250621209},\,\mathtt{-0.18822015672784},
\left.\mathtt{-0.57928211463475},\,\mathtt{-0.83834493032251}\right)\nonumber\\
\epsilon_7^{\mu}&=&\left(\right.\mathtt{0.021940447742366},\,\mathtt{-0.18167933678739},
\left.\mathtt{-0.55915153950179},\,\mathtt{-0.80921168905572}\right)\nonumber\\
\epsilon_8^{\mu}&=&\left(\right.\mathtt{0.29376006988772},\,\mathtt{-0.18931059303850},
\left.\mathtt{-0.58263813273021},\,\mathtt{-0.84320180521184}\right)\nonumber\\
\epsilon_9^{\mu}&=&\left(\right.\mathtt{2.9194377174610},\,\mathtt{0.56051931235228},
\left.\mathtt{1.7251011698101},\,\mathtt{2.4965898022169}\right)\nonumber\\
\eea
For color degree of freedom, we choose $\left\{a_i\right\}=\left\{1,1,1,2,2,2,3,3,3\right\}$. The full amplitude evaluates to
\be
\mathcal{A}_9^{\mathrm{tree}}\left(1^1_L,2^1_L,3^1_L,4^2_L,5^2_L,6^2_L,7^3_L,8^3_L,9^3_L\right)=-i\mathtt{8.941784390273400}\,\times 10^{-18}~\mathrm{GeV^{-5}}
\ee
The second benchmark configuration has both transverse and longitudinal polarizations $\left\{h_i\right\}=\left\{+,-,+,-,+,L,L,L,L\right\}$
\bea
\epsilon_1^{\mu}&=&\left(\right.\mathtt{0.0},\,\mathtt{-0.70710678118655},\left.\mathtt{0.70710678118655}i,\,\mathtt{0.0}\right)\nonumber\\
\epsilon_2^{\mu}&=&\left(\right.\mathtt{0.0},\,\mathtt{0.70710678118655},\left.\mathtt{-0.70710678118655}i,\,\mathtt{0.0}\right)\nonumber\\
\epsilon_3^{\mu}&=&\left(\right.\mathtt{0.0},\,\mathtt{0.17677668390637-0.67249851605560}i,\nonumber\\
&&\left.\mathtt{0.54406272447999+0.21850799963164}i,\,\mathtt{-0.41562694313348}\right)\nonumber\\
\epsilon_4^{\mu}&=&\left(\right.\mathtt{0.0},\,\mathtt{0.17677668390637+0.67249851605560}i,\nonumber\\
&&\left.\mathtt{0.54406272447999-0.21850799963164}i,\,\mathtt{-0.41562694313348}\right)\nonumber\\
\epsilon_5^{\mu}&=&\left(\right.\mathtt{0.0},\,\mathtt{0.17677668390637-0.67249851605560}i,\nonumber\\
&&\left.\mathtt{0.54406272447999+0.21850799963164}i,\,\mathtt{-0.41562694313348}\right)\nonumber\\
\epsilon_6^{\mu}&=&\left(\right.\mathtt{0.27169250621209},\,\mathtt{-0.18822015672784},
\left.\mathtt{-0.57928211463475},\,\mathtt{-0.83834493032251}\right)\nonumber\\
\epsilon_7^{\mu}&=&\left(\right.\mathtt{0.021940447742366},\,\mathtt{-0.18167933678739},
\left.\mathtt{-0.55915153950179},\,\mathtt{-0.80921168905572}\right)\nonumber\\
\epsilon_8^{\mu}&=&\left(\right.\mathtt{0.29376006988772},\,\mathtt{-0.18931059303850},
\left.\mathtt{-0.58263813273021},\,\mathtt{-0.84320180521184}\right)\nonumber\\
\epsilon_9^{\mu}&=&\left(\right.\mathtt{2.9194377174610},\,\mathtt{0.56051931235228},
\left.\mathtt{1.7251011698101},\,\mathtt{2.4965898022169}\right)\nonumber\\
\eea
Again we choose $\left\{a_i\right\}=\left\{1,1,1,2,2,2,3,3,3\right\}$. The full amplitude evaluates to
\bea
&&\mathcal{A}_9^{\mathrm{tree}}\left(1^1_+,2^1_-,3^1_+,4^2_-,5^2_+,6^2_L,7^3_L,8^3_L,9^3_L\right)\nonumber\\
&&=\left(\mathtt{-1.621175976214353}+\mathtt{2.231357479296454}i\right)\,\times 10^{-17}~\mathrm{GeV^{-5}}
\eea
By contrast, numerical implementation based on Feynman diagrams is too inefficient to yield a result for 9-$W$ scattering in a sensible amount of time.

\newpage

\bibliographystyle{JHEP}
\bibliography{color}{}

\end{document}